\documentclass[journal]{IEEEtran}

\renewcommand{\t}{^{\mbox{\tiny {T}}}}

\newcommand{\upfrown}[1]{
	\mathord{\buildrel{\lower3pt\hbox{$\scriptscriptstyle\frown$}} 
		\over #1}
}

\newcommand{\blue}{\color{blue}}
\newcommand{\eproof}{\hfill\rule{2mm}{2mm}}

\newcommand{\bstate}{\medskip\begin{state} }
	\newcommand{\estate}{ \hfill  \rule{1mm}{2mm}\medskip\end{state}}

\newcommand{\bass}{\medskip\begin{ass} }
	\newcommand{\eass}{ \hfill  \rule{1mm}{2mm}\medskip\end{ass}}

\newcommand{\brem}{\medskip \begin{remark}  }
	\newcommand{\erem}{\hfill \rule{1mm}{2mm}\medskip
\end{remark} }
\newcommand{\bthm}{\medskip\begin{theorem}  }
	\newcommand{\ethm}{ \hfill  \rule{1mm}{2mm} \medskip
\end{theorem} }
\newcommand{\blem}{\medskip\begin{lemma}  }
	\newcommand{\elem}{ \hfill \rule{1mm}{2mm}\medskip
\end{lemma} }
\newcommand{\bcorollary}{\medskip\begin{corollary}  }
	\newcommand{\ecorollary}{  \hfill \rule{1mm}{2mm}\medskip
\end{corollary} }
\newcommand{\bdefn}{\medskip\begin{definition}}
	\newcommand{\edefn}{  \hfill \rule{1mm}{2mm}\medskip
\end{definition} }
\newcommand{\bproposition}{\medskip\begin{proposition} }
	\newcommand{\eproposition}{\hfill \rule{1mm}{2mm}\medskip
\end{proposition} }
\newcommand{\bexample}{\medskip\begin{example} \rm}
	\newcommand{\eexample}{ \hfill \rule{1mm}{2mm}\medskip
\end{example} }

\newcommand{\bcon}{\medskip\begin{condition} \rm}
	\newcommand{\econ}{ \hfill \rule{1mm}{2mm}\medskip
\end{condition} }

\newcommand{\prooflater}[1]{\noindent{\bf Proof of #1: }}

\newtheorem{theorem}{\bf Theorem}[section]
\newtheorem{ass}{\bf Assumption}[section]
\newtheorem{lemma}{\bf Lemma}[section]
\newtheorem{definition}{\bf Definition}[section]
\newtheorem{remark}{\bf Remark}[section]
\newtheorem{corollary}{\bf Corollary}[section]
\newtheorem{proposition}{\bf Proposition}[section]
\newtheorem{example}{\bf Example}[section]
\newtheorem{condition}{\bf Condition}[section]
\newtheorem{state}{\bf Assumption}[section]

\usepackage{color}
\usepackage{amsmath,amsfonts}
\usepackage{subfigure}

\usepackage{verbatim}
\usepackage{graphicx}
\usepackage{ulem}

\usepackage{xcolor,soul,framed} 

\colorlet{shadecolor}{yellow}

\begin{document}
\bstctlcite{IEEEexample:BSTcontrol}
    \title{An Improved Fractional-Order Active Disturbance Rejection Control: Performance Analysis  and  Experiment Verification}
  \author{Bolin Li, 
       Lijun Zhu,

\thanks{Bolin Li is with Key Laboratory of Imaging Processing and Intelligence Control, School of Artificial Intelligence and Automation, Huazhong University of Science and Technology, Wuhan 430074, China.}
\thanks{Lijun Zhu is with Key Laboratory of Imaging Processing and Intelligence Control, School of Artificial Intelligence and Automation, Huazhong University of Science and Technology, Wuhan 430074, China.}}

\markboth{An Improved Fractional-Order Active Disturbance Rejection Control: Performance Analysis  and  Experiment Verification
}{Bolin Li \MakeLowercase{\textit{et al.}}: An Improved Fractional-Order Active Disturbance Rejection Control: Performance Analysis  and  Experiment Verification}

\maketitle

\begin{abstract}
This paper presents an improved active disturbance rejection control scheme (IFO-ADRC) with an improved fractional-order extended state observer (IFO-ESO). 
The structural information of the system is utilized  in  IFO-ESO rather than  buried as in the typical fractional-order extended state observer (FO-ESO) and help significantly improve the performance of IFO-ESO and closed-loop system. Compared with the integer-order active disturbance rejection controller (IO-ADRC), the auxiliary tracking controller of IFO-ADRC has a simpler form and fewer parameters need to be tuned. Frequency-domain analysis shows that IFO-ESO has better performance over the larger frequency band than FO-ESO, and time-domain simulation  shows that IFO-ADRC has better transient performance and is more robust against the parameter variations than traditional fractional-order active disturbance rejection controller (FO-ADRC) and IO-ADRC. The IFO-ADRC is applied to permanent magnet synchronous motor (PMSM) servo control system and demonstrates its  capability in the real-world application.
\end{abstract}


%
\IEEEpeerreviewmaketitle

\section{Introduction}
\IEEEPARstart{F}{ractional} calculus has been applied to different fields in recent years \cite{chen2021fractional,liu2018guaranteed,pu2016fractional,tzounas2020theory}. With a deep understanding of the system, there is a growing need for fractional-order control and modelling \cite{luo2009fractional,mishra2013fractional}. Some real-world phenomena demonstrate fractional-order characteristics \cite{chen2009fractional,zheng2016fractional}. Permanent magnet synchronous motor \cite{zheng2016fractional}, gas-turbine \cite{nataraj2010computation}, and  heating–furnace \cite{podlubny1997fractional} can be identified as fractional-order models. On the other hand, the fractional-order controller has the potential to achieve better and more robust control performance than the integer-order controller \cite{zheng2021synthesis}. Many fractional-order controllers have been proposed, including fractional-order sliding mode controller \cite{zaihidee2019application,ren2019fractional}, fractional-order intelligent PID controller \cite{gomaa2019novel}, fractional-order PID controller \cite{li2009fractional}, and so on.

The active disturbance rejection control (ADRC) was proposed in  \cite{2002From}  as an alternative paradigm for control system design. The core idea of ADRC is to improve the robustness of the system  using extended state observer (ESO). ESO, an important part   ADRC, can estimate the total disturbances, including internal disturbance caused by system uncertainties and external disturbance. Gao \cite{gao2006scaling} proposed the linear ADRC to broaden the application range of ADRC by linearizing ESO (IO-ESO) and feedback control law.  Applying IO-ESO, a $m$-order integer-order system can be approximately converted into $m$ unit gain integer-order integrators in series. According to the design idea of the linear IO-ADRC, a linear feedback control law of $m$ parameters need to be designed to realize the stable closed-loop system. Furthermore,   Gao \cite{gao2016active} proposed an ADRC structure involving the fractional-order tracking differentiator, the fractional-order PID controller and the fractional-order extended state observer for nonlinear fractional-order systems. The stability region of the fractional-order system can be larger than that of the integer-order system in the complex plane \cite{kumar2017stability}.  FO-ADRC can provide a possibility to realize closed-loop stability with a simpler controller. In \cite{chen2021fractional}, Chen et al. applied FO-ESO to approximately convert a typical second-order system into a cascaded fractional-order integrator and the stable closed-loop system can be realized using merely a simple proportional controller. For a fractional-order system, it is natural to use a fractional-order controller to achieve closed-loop stability \cite{choudhary2014stability,trivedi2020design}. A FO-ADRC based on FO-ESO was proposed by Li \cite{li2016fractional} to approximately convert the fractional-order system to a cascaded fractional-order integrator.
The IO-ADRC was proved by Li \cite{li2013active} to estimate the disturbances in fractional-order systems, which considers the fractional-order dynamics as a common disturbance.

	An integer-order system can be approximately converted to a fractional-order system (a cascaded fractional-order integrator) by using FO-ESO and then an auxiliary tracking controller is designed to 
	meet the actual needs \cite{chen2021fractional}. However, the
	satisfactory approximation can only be made 
	 in the low-frequency band, making it difficult to ensure the system's robustness when the control system bandwidth is large.
This paper is along the research line of proposing a new type of fractional-order ESO structure and then fractional-order ADRC. The main contributions of the paper can be summarized as follows. First, the new type of ESO, called IFO-ESO, will utilize structural information in the total disturbance given in FO-ESO. IFO-ESO will be used to compensate for the system uncertainties and external disturbance. The compensated system is approximately converted into a fractional-order integrator. It will be shown that 
the approximation performance is improved by IFO-ESO  over FO-ESO,  making the system more robust to  system uncertainties and external disturbances. The theoretical analysis is verified by the control experiment on the PMSM servo speed control system. Second, compared with IO-ADRC, the auxiliary tracking controller of IFO-ADRC has a simpler form and fewer parameters need to be tuned. Thirdly, the stability criteria of the IFO-ESO and the IFO-ADRC closed-loop system are given when the disturbance is assumed to be bounded. Unlike the stability analysis of the ESO and the closed-loop system in \cite{chen2021fractional} and \cite{li2016fractional}, the stability criteria of the high-order ESO and high-order ADRC closed-loop system in this paper are given.

This paper is organized as follows. The structures of IFO-ESO and IFO-ADRC are proposed in Section \uppercase\expandafter{\romannumeral2}. The BIBO stability criteria of the IFO-ESO and the closed-loop system are given in Section \uppercase\expandafter{\romannumeral3}. The performance analysis of the IFO-ESO in the frequency-domain is shown in Section \uppercase\expandafter{\romannumeral4}. Section \uppercase\expandafter{\romannumeral5} presents the time-domain simulation results, followed by the experimental results on PMSM servo system  in \uppercase\expandafter{\romannumeral6}. The paper is concluded in Section \uppercase\expandafter{\romannumeral7}.

\textbf{Notations}. $I_n$ is an $n\times n$ identity matrix and $0_{n\times m}$ is an  zero vector matrix with size specified by the subscript.

\section{An improved Structure of Active disturbance rejection controls \label{sec:ADRC}}
In this paper, we consider an integer-order linear system as follows:
\begin{equation}
G(s) = \frac{{Y(s)}}{{U(s)}} = \frac{b}{{{s^{m }} + \sum\limits_{i = 1}^{m - 1} {{a_i}{s^{i }}}  + {a_0}}}
\label{eq_6}
\end{equation}
where $s$ is Laplace operator, $a_i$, $a_0$ and $b$ are real numbers, $m$ and $i$ are positive integers with $m$ representing the maximum order of the system. 
The differential equation form of system (\ref{eq_6}) with the external disturbance, denoted by $d$, is
\begin{equation}
{y^{(m )}} =  - \mathop \sum \limits_{i = 1}^{m - 1} {a_i}{y^{(i)}} - {a_0}y + bu + d
\label{eq_7}
\end{equation}
Caputo derivative is adopted as the fractional-order derivative method in this paper and described as follows,
\begin{equation}
{f^{(\gamma )}}(t) = {}_0^CD_t^\gamma f(t) = \frac{1}{{\Gamma ({m_0} - \gamma )}}\int_0^t {\frac{{{f^{({m_0})}}(\tau )}}{{{{(t - \tau )}^{\gamma  - {m_0} + 1}}}}d\tau } 
\label{eq:sys}
\end{equation}
where $m_0$ is an integer satisfying $m_0 - 1 < \gamma  < m_0$ and $\gamma$ is a real number, $\Gamma ( \bullet )$ is Euler's gammafunction.  From the adopted definition, one can see that ${}_0^CD_t^\gamma f{(t)^{(\vartheta)}} = {}_0^CD_t^{\gamma  + \vartheta }f(t)$ where $\vartheta$ is a positive number \cite{li2011riemann}.

Let $\bar \gamma$ be  a fractional number satisfying $m-1<\bar \gamma<m$, $n$ be  a positive integer number satisfying $\bar\gamma < n <\frac{\bar\gamma}{m-\bar\gamma}$, and $\gamma=\frac{\bar\gamma}{n}$ be a fractional number. As a result, $n\gamma < m < (n+1)\gamma$. 
  Define the quantity $q({y^{(n\gamma )}},{y^{(m)}},t) = {y^{(n\gamma)}} - {y^{(m)}}$. Equation (\ref{eq_7}) can be rewritten as follows,
\begin{align}
{y^{(n\gamma )}}  = &{y^{(n\gamma )}} - {y^{(m)}} - \sum\limits_{i = 1}^{m - 1} {{a_i}} {y^{(i )}} - {a_0}y + bu + d \nonumber \\
=& {y^{(n\gamma )}} - {y^{(m)}} - \sum\limits_{i = 1}^{m - 1} {{a_i}} {y^{(i)}} - {a_0}y + (b - {b_0})u  \nonumber \\&+ {b_0}u + d \nonumber\\
= &q({y^{(n\gamma )}},{y^{(m )}},t) + {f_{ifo}}({y^{(1)} },{y^{(2) }}, \cdots ,{y^{(m - 1) }},y,u,t)\nonumber  \\&+ {b_0}u{\mkern 1mu}
\label{eq_8}
\end{align}
Note that $f_{ifo}=- \sum\limits_{i = 1}^{m - 1} {{a_i}} {y^{(i )}} - {a_0}y + (b - {b_0})u+d$  can be regarded as the total disturbance where the term $- \sum\limits_{i = 1}^{m - 1} {{a_i}} {y^{(i )}} - {a_0}y + (b - {b_0})u$ is the internal disturbance due to  uncertain parameters and $d$ is the external disturbance. The aim of this paper is to design $u$ such that the system output tracks a sufficiently smooth reference trajectory $r$ and the ultimate tracking error stays in the neighborhood of the origin,  i.e., $\lim_{t\rightarrow \infty} \|y(t)-r(t)\|<\epsilon$, when   the reference signal and its derivatives, i.e., $r$, $ \dot r,\ddot r, \cdots {r^{(m-1)}},{r^{(n\gamma-m+1)}}$, are  bounded.  

As inspired by FO-ADRC in \cite{chen2021fractional}, we propose the tracking controller as follows
\begin{equation}
u = \frac{u_0- \upfrown{q}  - \upfrown{f}_{ifo}}{b_0}
\label{eq_8_}
\end{equation}
where 
$\upfrown{q}$ and $\upfrown{f}_{ifo}$ are estimators for signals $q$ and $f_{ifo}$, respectively, and
$u_0$ is the auxiliary tracking controller to be specified later. 
If signals $q$ and $f_{ifo}$ are approximately estimated by  $\upfrown{q}$ and $\upfrown{f}_{ifo}$, the closed-loop system composed of (\ref{eq_8}) and (\ref{eq_8_}) can be approximately converted into a  fractional-order integrator  as follows
\begin{equation}
{y^{(n\gamma)}} = {u_0} + (f_{ifo} - \upfrown{ f}_{ifo}) + (q - \upfrown{ q} ) \approx {u_0}.
\label{eq_14}
\end{equation}

As will be explained in Remark \ref{rem:IO-ADRC},  a big advantage of IFO-ADRC and FO-ADRC is that a simpler auxiliary controller can be designed and easily tuned.
\brem
For FO-ADRC in \cite{chen2021fractional}, equation (\ref{eq_7}) is written as
\begin{align}
{y^{(n\gamma )}}   
=& {f_{fo}}({y^{(1)} },{y^{(2) }}, \cdots ,{y^{(m - 1) }},{y^{(n\gamma )}},{y^{(m )}},y,u,t)\nonumber  \\&+ {b_0}u{\mkern 1mu}
\end{align}
where the term  $f_{fo}$ = ${y^{(n\gamma )}} - {y^{(m)}} - \sum\limits_{i = 1}^{m - 1} {{a_i}} {y^{(i )}} - {a_0}y + (b - {b_0})u$. 
In comparison to FO-ADRC,  IFO-ADRC will separate the structurally certain term $q$ from $f_{fo}$. It is  explicitly included in (\ref{eq_8}) and estimated in a new type of ESO introduced later.  As will be demonstrated in Section \ref{sec:PA} and \ref{sec:TDS}, the structural certainty of the term $q$ can help significantly improve the performance of ESO and the closed-loop system.
\erem

Let ${x_1} = y,{x_2} = {y^{(\gamma)}}, \cdots ,{x_{n}} = {y^{((n-1)\gamma)}}, {x_{n+1}} = {f_{ifo}}, {h_{ifo}} = {{f}^{(\gamma)}_{ifo}}({y^{(1)} },{y^{(2) }}, \cdots ,{y^{(m - 1) }},y,u,t)$ where $x_1, x_2, \cdots, {x_{n}}$ represent system states and ${x_{n+1}}$ is an extended state, the state-space representation of (\ref{eq_8}) is given as follows:
\begin{equation}
\left\{ {\begin{array}{*{20}{l}}
	{x^{(\gamma)}  = Ax + Bu + E{h_{ifo}} + Fq}\\
	{y = Cx}
	\end{array}} \right.
\label{eq_9}
\end{equation}
where 
\begin{gather}
x = {\left[{\begin{array}{*{20}{c}}
		{{x_1}},{{x_2}}, \cdots ,{{x_n}},{{x_{n + 1}}}
		\end{array}} \right]\t}, \;A =\left[\begin{array}{cc}
0_{n\times1} & I_{n} \nonumber\\
0 & 0_{1\times n}
\end{array}\right] \nonumber\\
B = {\left[{\begin{array}{*{20}{c}}
		0,0, \cdots ,{{b_0}},0
		\end{array}} \right]\t},
C = \left[ {\begin{array}{*{20}{c}}
	1,0, \cdots ,0,0
	\end{array}} \right],\nonumber\\
E = {[\begin{array}{*{20}{c}}
	0& \cdots &0&1
	\end{array}]\t},
F = {[\begin{array}{*{20}{c}}
	0& \cdots & 0&1&0
	\end{array}]\t}.
\end{gather}
Then,  IFO-ESO is designed to estimate  ${x_1},{x_2} \cdots {x_{n}},{x_{n+1}}$ as follows:
\begin{gather}
z^{(\gamma)} = Az + Bu + F\mathord{\buildrel{\lower3pt\hbox{$\scriptscriptstyle\frown$}} 
	\over q}  + L(y - \upfrown{y}) \nonumber\\
\upfrown{y} = Cz
\label{eq_10}
\end{gather}
where
\begin{gather} 
z = \left[{\begin{array}{*{20}{c}}
	{{z_1}},{{z_2}}, \cdots ,{{z_n}},{{z_{n + 1}}}
	\end{array}} \right]\t \nonumber \\
L = \left[ {\begin{array}{*{20}{c}}
	{{\beta _1}},{{\beta _2}}, \cdots ,{{\beta _n}},{{\beta _{n + 1}}}
	\end{array}} \right]\t \nonumber \\
{\mathord{\buildrel{\lower3pt\hbox{$\scriptscriptstyle\frown$}} 
		\over q}  = ({z^{(\gamma)}_{n}} - {z_{n}}^{(m-n\gamma+\gamma)})}\label{eq:ABC}
\end{gather}
Note that $L$ are extended state observer gains. $z_1, z_2,  \cdots, z_{n}, z_{n+1}$ are the estimation of the state $x_1, x_2,  \cdots, x_{n}, x_{n+1}$, respectively and $b_0$ is the nominal value of $b$. Let $\upfrown{f}_{ifo}$ in (\ref{eq_8_}) be $
\upfrown{f}_{ifo}	= {z_{n+1}}.$

With IFO-ESO, the tracking task can be fulfilled with 
the auxiliary tracking controller $u_0 $ in (\ref{eq_8_}), which  is designed as follows,
\begin{gather}
{u_0} = {k_p}({r} - {z_1}) + {k_{{d_1}}}({\dot{r}} - {\dot {z_1}}) +  \ldots \nonumber \\
+ {k_{{d_{m - 2}}}}({r}^{(m-2)} - {z_1}^{(m-2)}) + {r^{(n\gamma)}}	
\label{eq_12}
\end{gather}
where 
  ${k_p},{k_{d_1}}, \cdots ,{k_{{d_{m-2}}}}$ are the parameters of the feedback control law. The closed-loop system composed of (\ref{eq_7}),  (\ref{eq_8_}) and (\ref{eq_12}), is called IFO-ADRC system.

\brem \label{rem:IO-ADRC} For IO-ADRC in \cite{gao2006scaling}, the (\ref{eq_7}) is written as
\begin{equation}
{y^{(m)}} 
= {f_{io}}({y^{(1)} },{y^{(2) }}, \cdots ,{y^{(m - 1) }},y,u,t)\nonumber  + {b_0}u{\mkern 1mu}
\end{equation}
where  $f_{io}$ = $- \sum\limits_{i = 1}^{m - 1} {{a_i}} {y^{(i )}} - {a_0}y + (b - {b_0})u$. Any part of controller similar to (5) where the auxiliary tracking controller $u_0$ is designed as
\begin{gather}
{u_0} = {k_p}({r} - {z_1}) + {k_{{d_1}}}({\dot{r}} - {\dot {z_1}}) +  \ldots \nonumber \\
+ {k_{{d_{m - 1}}}}({r}^{(m-1)} - {z_1}^{(m-1)}) + r^{(m)},
\label{eq_13}
\end{gather}
Note that the number of parameters for the auxiliary tracking controller of IFO-ADRC is less than that of IO-ADRC.
 In other words, we can achieve the trajectory tracking for an integer-order plant with a simpler auxiliary controller $u_0$ process,  which simplifies the parameter tuning.
As will be demonstrated in Section \ref{sec:TDS}, a P auxiliary tracking controller can be adopted to achieve the  trajectory tracking of a second-order integer-order plant, and a PD auxiliary tracking controller can be adopted for a third-order integer-order plant.
Section \ref{sec:TDS} will show the system transient performance of IFO-ADRC is  better than that of  IO-ADRC.
\erem

\section{Stablilty analysis of IFO-ADRC}
In this section, the stability criteria for the ESO and the IFO-ADRC system are provided.
Let the observer error be
\begin{equation} \label{eq:ei}
{e_i} = {x_i} - {z_i}, i=1,\cdots,n+1.
\end{equation}
From (\ref{eq_9}) and (\ref{eq_10}), the  equation of the extended state observer error can be written as
\begin{equation}
e^{(\gamma)} = Ae-Le_1+F({e_n}^{(\gamma )} - {e_n}^{(m  - n\gamma  + \gamma )})
\label{eq_30_}
\end{equation}
where $e = {\left[{\begin{array}{*{20}{c}}
		{{e_1}},{{e_2}}, \cdots ,{{e_n}},{{e_{n + 1}}}
		\end{array}} \right]\t}$.
The characteristic matrix of (\ref{eq_30_}) is \cite{deng2007stability}:
\begin{equation}
\lambda (s) = \left[ {\begin{array}{*{20}{c}}
	{{s^\gamma } + {\beta _1}}&{ - 1}&0& \cdots &0&0\\
	{{\beta _2}}&{{s^\gamma }}&{ - 1}& \cdots &0&0\\
	\vdots & \vdots & \vdots & \vdots & \vdots & \vdots \\
	{{\beta _n}}&0&0& \cdots &{{s^\chi }}&{ - 1}\\
	{{\beta _{n + 1}}}&0&0& \cdots &0&{{s^\gamma }}
	\end{array}} \right]
\label{eq_31_}
\end{equation}
From (\ref{eq_31_}), the characteristic polynomial of the system (\ref{eq_30_}) can be obtained:
\begin{equation}
\lambda (s) = {s^{\nu+\gamma} }({s^{n\gamma }} + \sum\limits_{i = 1}^{n - 1} {{\beta _i}} {s^{(n - i)\gamma }}) + {\beta _{n}}{s^\gamma } + {\beta _{n+1}}
\label{eq_32_}
\end{equation}
where $\nu = m-n\gamma$ and $0<\nu<\gamma$.
Then,
Theorem \ref{th_IFO-ESO} will present the  bounded-input bounded-output (BIBO) stability of the error system (\ref{eq_30_}), when $h_{ifo}$ is bounded. If the boundedness of the external disturbance $d$ rather than $h_{ifo}$ is assumed, Theorem \ref{th_closed-loop} will give the condition of BIBO  stability of the closed-loop system in terms of roots of a polynomial. In this case, the tracking error converges into a neighborhood of the origin. 
A special case of Theorem \ref{th_closed-loop} when $m=n=2$ is elaborated in Proposition \ref{prop:stable}, showing that when the observer gain is selected sufficiently large, the system is BIBO.
The proofs of both theorems and the proposition are given in the Appendix.
\bthm Consider the error dynamics of IFO-ESO (\ref{eq_30_}). Let   $\omega_o > 0$ and ${\beta _i} = C_{n + 1}^i{\omega _o}^i$ for $i = 1,2, \cdots ,n+1$. If $h_{ifo}$ is bounded,  then the IFO-ESO is  BIBO stable, regarding  $h_{ifo}$ as the input and $e_1$ as the output.
\label{th_IFO-ESO}
\ethm

\bthm  Consider the IFO-ADRC closed-loop system  composed of (\ref{eq_7}),  (\ref{eq_8_}) and (\ref{eq_12}). Let $p_1$, $p_2$, $q_1$, and $q_3$  positive prime such that $\nu = \frac{p_1}{q_1}$ and $\gamma = \frac{p_2}{q_2}$. Define a polynomial 
\begin{gather}
P(w) = ({w^{{p_2}{q_1}}}\sum\limits_{i = 0}^{m - 1} {{a_i}{w^{i{q_1}{q_2}}}} )({k_p} + \sum\limits_{i = 1}^{m - 2} {{k_{{d_i}}}{w^{i{q_1}{q_2}}}} \nonumber \\
+ {w^{n{p_2}{q_1}}} + \sum\limits_{i = 1}^n {{\beta _i}{w^{(n - i){p_2}{q_1}}}} ) + ({w^{n{p_2}{q_1}}} + {k_p}\nonumber\\
+ \sum\limits_{i = 1}^{m - 2} {{k_{{d_i}}}{w^{i{q_1}{q_2}}}} )({w^{{p_1}{q_2} + {p_2}{q_1}}}({w^{n{p_2}{q_1}}} + \sum\limits_{i = 1}^{n - 1} {{\beta _i}} {w^{(n - i){p_2}{q_1}}})\nonumber\\
+ {\beta _n}{w^{{p_2}{q_1}}} + {\beta _{n + 1}})
\label{eq-19}
\end{gather}
 If $b=b_0$, ${{k_p},{k_d}_{_1}, \cdots , {k_d}_{_{m - 2}}}$ and $\beta_1, \cdots, \beta_{n+1}$ are selected such that   all the roots of (\ref{eq-19}) are located in $|\mbox{arg}({w_i})| > \frac{\pi }{{2{q_1}{q_2}}}$, then the IFO-ADRC closed-loop system is BIBO stable,
 Moreover, the tracking error $r(t)-y(t)$ converges to a small neighborhood of the origin as $t\rightarrow \infty$.
\label{th_closed-loop}
\ethm

	\bproposition	\label{prop:stable}
	Consider the IFO-ADRC closed-loop system  composed of (\ref{eq_7}),  (\ref{eq_8_}) and (\ref{eq_12}) with
	$m=n=2$. Suppose the plant 
	(\ref{eq_7}) is stable or marginally stable, i.e., $a_1 \ge 0$ and $a_0 \ge 0$.
	Let ${\beta _i} = C_{n + 1}^i{\omega _o}^i$ for $i = 1,2,3$, $k_p>0$. Then, there always exists a constant $\omega_o > 0$, such that the closed-loop system is BIBO stable.
	 Moreover, the tracking error $r(t)-y(t)$ converges to a small neighborhood of the origin as $t\rightarrow \infty$.
	\eproposition
\section{Performance Analysis of IFO-ESO in Frequency- domain \label{sec:PA}}
In this section, we will  compare the performance of the IFO-ESO proposed in Section \ref{sec:ADRC} with the FO-ESO proposed in   \cite{chen2021fractional}. 
The  FO-ESO   is  as follows
\begin{gather}
z^{(\gamma)} = A z +Bu+L(y-\upfrown{y})\nonumber\\
\upfrown{y} = C z
\label{eq_56}
\end{gather}
where  $z$, $A$, $B$, $C$ and $L$ are given in  (\ref{eq:ABC}).
Similar to the analysis in Section \ref{sec:ADRC}, the controller $u$ utilizing the estimation of the FO-ESO in \cite{chen2021fractional} is 
\begin{equation} \label{eq:FO-ADRC}
u = \frac{u_0  - \upfrown{f}_{fo}}{b_0}
\end{equation}
which is aimed to approximately convert the perturbed   system into a pure cascaded integrator shown as follows
\begin{equation}
{y^{(n\gamma)}} = {u_0} + (f_{fo} - \upfrown{ f}_{fo}) \approx {u_0}.\label{eq:FO-App}
\end{equation}

Note that the role of the ESO in the framework of ADRC including IO-ESO, FO-ESO and IFO-ESO  is  to estimate the uncertain dynamics and external disturbances  to improve the robustness of the system. 
If the IFO-ESO can perfectly estimate $\upfrown{q}$ and $\upfrown{f}_{ifo}$ or the FO-ESO  can perfectly estimate $\upfrown{f}_{fo}$, 
the IFO-ADRC in (\ref{eq_8_})  and the FO-ADRC in (\ref{eq:FO-ADRC})  can  convert the original system  into  a cascaded fractional-order integrator $1/s^{n\gamma}$ (looking from $u_0$ to $y$).
Therefore, we are motivated to use the model difference between $Y(s)/U_0(s)$ and $1/s^{n\gamma}$  to assess the performance of the two ESO.
We adopt mean square error between $Y(s)/U_0(s)$ and $1/s^{n\gamma}$  in the frequency-domain  to evaluate how difference  the two models  are. The mean square error  (MSE) of two linear model is defined as 
\begin{equation}
{e_o(\omega )} = |{\Delta _o(\omega)}{|^2}
\label{eq_43}
\end{equation}
where
\begin{equation}
{\Delta _o(\omega)} = 1 - (j\omega)^{n\gamma}Y(j\omega)/U_0(j\omega)
\end{equation}
The MSE  (\ref{eq_43}) was used in \cite{richardson1982parameter} for the model  identification where the problem is re-casted into an optimal problem of minimizing the model difference between the  identified and ideal model in terms of the MSE. Therefore, the MSE can be used to evaluate model difference in the frequency-domain.

As in  \cite{chen2021fractional}, for simplicity, we  consider the  second-order system as follow
\begin{equation} 
G(s) = \frac{Y(s)}{U(s)}=\frac{b_0}{{s(s + {a_o})}},
\label{eq_53_}
\end{equation}
where the external disturbance and system uncertainty are not considered.
%
In this case,  the  FO-ESO  in (\ref{eq_56}) is simplified 
with
\begin{gather}
z = {\left[{\begin{array}{*{20}{c}}
		{{z_1}},{{z_2}}, {{z_{3}}}
		\end{array}} \right]\t}, \;A =\left[\begin{array}{cc}
0_{3\times1} & I_{3} \\
0 & 0_{1\times 3}
\end{array}\right] \nonumber\\
B = {\left[{\begin{array}{*{20}{c}}
		0,{{b_0}},0
		\end{array}} \right]\t},
C = \left[ {\begin{array}{*{20}{c}}
	1,0,0
	\end{array}} \right]. \label{eq:FO-ESO3}
\end{gather}
Similarly, the IFO-ESO is 
\begin{gather} 
z^{(r)}=Az+Bu+E\upfrown{q} +L(y-\upfrown{y}),\nonumber\\
\upfrown{q}= ({z_{2}^{(\gamma )}} - z_{2}^{(2-\gamma)}),\nonumber\\
\upfrown{y}=Cz,\label{eq:IFO-ESO2}
\end{gather}
where $E=\left[ {\begin{array}{*{20}{c}}
	0,1 ,0
	\end{array}} \right]\t$.

Next, let us calculate $Y(s)/U_0(s)$ for FO-ADRC and IFO-ADRC.
For the fair comparison, we choose  the same group of  the observer gains ${\beta _1} = 3{\omega _o}$, ${\beta _2} = 3{\omega _o}^2$, ${\beta _3} = {\omega _o}^3$  for IFO-ESO and FO-ESO. 
%
%
%
%
For the IFO-ESO, all the parameters meet the conditions of Theorem \ref{th_IFO-ESO}, and thus the dynamics of the estimation  error for IFO-ESO  is asymptotically stable (because the disturbance and uncertainties do not exist). Conducting the Laplace transform on the both sides of  (\ref{eq:IFO-ESO2})  gives
\begin{gather}
{Z_1}(s) = \frac{{(3{\omega _o}{s^2} + 3{\omega _o}^2{s^\gamma } + {\omega _o}^3)Y(s)}}{{{s^{2 + \gamma }} + 3{s^2}{\omega _o} + 3{s^\gamma }{\omega _o}^2 + {\omega _o}^3}}  \nonumber\\+ \frac{{{b_0}{s^\gamma }U(s)}}{{{s^{2 + \gamma }} + 3{s^2}{\omega _o} + 3{s^\gamma }{\omega _o}^2 + {\omega _o}^3}}\nonumber \\
{Z_2}(s) = \frac{{(3{\omega _o}^2{s^{2\gamma }} + {\omega _o}^3{s^\gamma })Y(s)}}{{{s^{2 + \gamma }} + 3{s^2}{\omega _o} + 3{s^\gamma }{\omega _o}^2 + {\omega _o}^3}}\nonumber \\+ \frac{{{b_0}({s^{2\gamma }} + 3{\omega _o}{s^\gamma })U(s)}}{{{s^{2 + \gamma }} + 3{s^2}{\omega _o} + 3{s^\gamma }{\omega _o}^2 + {\omega _o}^3}}\nonumber\\
{Z_3}(s) = \frac{{{\omega _o}^3{s^2}Y(s)}}{{{s^{2 + \gamma }} + 3{s^2}{\omega _o} + 3{s^\gamma }{\omega _o}^2 + {\omega _o}^3}} \nonumber \\ - \frac{{{b_0}{\omega _o}^3U(s)}}{{{s^{2 + \gamma }} + 3{s^2}{\omega _o} + 3{s^\gamma }{\omega _o}^2 + {\omega _o}^3}}\nonumber\\
{Q(s) = ({s^\gamma } - {s^{2 - \gamma }}){Z_2}(s)}
\label{eq_57}
\end{gather}
where $Z_1(s)$, $Z_2(s)$, $Y(s)$, $U(s)$ and $Q(s)$ are the Laplace transforms of signals $z_1$, $z_2$, $y$, and $u$, and the quantity $q$, respectively.
Conducting the Laplace transform on the both sides of  (\ref{eq_8_}) and substituting the result and (\ref{eq_57}) into (\ref{eq_53_}) obtain
\begin{equation}
{P_{ifo}}(s) = \frac{{Y(s)}}{{{U_0}({\rm{s}})}} = \frac{{{N_1}}}{{{D_1}}}
\label{eq_57_}
\end{equation}
where 
\begin{align}
{N_1} =& b\left( {{s^{2 + \gamma }} + 3{s^2}{\omega _o} + 3{s^\gamma }{\omega _o}^2 + {\omega _o}^3} \right) \nonumber\\
{D_1} = &b{\omega _o}^2{s^\gamma }\left( { - 3{s^2} + 3{s^{2\gamma }} + {s^\gamma }{\omega _o}} \right)\nonumber\\
&+ {a_o}{b_0}{s^{1 + \gamma }}\left( {{s^{2\gamma }} + 3{s^\gamma }{\omega _o} + 3{\omega _o}^2} \right)\nonumber\\
&+ {b_0}{s^{2 + \gamma }}\left( {{s^{2\gamma }} + 3{s^\gamma }{\omega _o} + 3{\omega _o}^2} \right)
\end{align}

Similarly, the Laplace transform of the both side of  FO-ESO  (\ref{eq_56}) gives
\begin{gather}
{Z_1}(s) = \frac{{(3{\omega _o}{s^{2\gamma }} + 3{\omega _o}^2{s^\gamma } + {\omega _o}^3)Y(s)}}{{{s^{3\gamma }} + 3{s^{2\gamma }}{\omega _o} + 3{s^\gamma }{\omega _o}^2 + {\omega _o}^3}} \nonumber\\+ \frac{{{b_0}{s^\gamma }U(s)}}{{s^{3\gamma } + 3{s^{2\gamma }}{\omega _o} + 3{s^\gamma }{\omega _o}^2 + {\omega _o}^3}}\nonumber\\
{Z_2}(s) = \frac{{(3{\omega _o}^2{s^{2\gamma }} + {\omega _o}^3{s^\gamma })Y(s)}}{{{s^{3\gamma }} + 3{s^{2\gamma }}{\omega _o} + 3{s^\gamma }{\omega _o}^2 + {\omega _o}^3}} \nonumber\\
+ \frac{{{b_0}({s^{2\gamma }} + 3{\omega _o}{s^\gamma })U(s)}}{{{s^{3\gamma }} + 3{s^{2\gamma }}{\omega _o} + 3{s^\gamma }{\omega _o}^2 + {\omega _o}^3}}\nonumber\\
{Z_3}(s) = \frac{{{\omega _o}^3{s^2}Y(s)}}{{{s^{3\gamma }} + 3{s^{2\gamma }}{\omega _o} + 3{s^\gamma }{\omega _o}^2 + {\omega _o}^3}} \nonumber\\
- \frac{{{b_0}{\omega _o}^3U(s)}}{{{s^{3\gamma }} + 3{s^{2\gamma }}{\omega _o} + 3{s^\gamma }{\omega _o}^2 + {\omega _o}^3}}
\label{eq_57_1}
\end{gather}
Conducting the Laplace transform on the both sides of   (\ref{eq:FO-ADRC}) and substituting the result and (\ref{eq_57_1}) into (\ref{eq_53_}) obtain
\begin{equation}
{P_{fo}}(s) = \frac{{Y(s)}}{{{U_0}(s)}} = \frac{{{N_2}}}{{{D_2}}}
\label{eq_60}
\end{equation}
where
\begin{equation}
\left\{ \begin{array}{l}
{N_2} = b{({s^\gamma } + {\omega _o})^3}\\
{D_2} = bs({a_o} + s){({s^\gamma } + {\omega _o})^3}\\
\quad \quad  + (b{s^{2\gamma }} - {b_0}{s^2} - {a_o}{b_0}s){\omega _o}^3
\end{array} \right.
\end{equation}

The model difference  between the fractional-order integrator $(1/s^{2\gamma})$ and $P_{o}$ ($P_{ifo}$ or $P_{fo}$) at $\omega$ can be expressed as:
\begin{equation}
{\Delta _o} = 1 - (j\omega)^{2\gamma}{P_o}(j\omega )
\end{equation}
Then, the mean-square error  between the  fractional-order integrator  and $P_{ifo}$ can be expressed as follows:
\begin{equation}
{\Delta_{ifo}} = \frac{{{N_3}}}{{{D_3}}}
\label{eq_44}
\end{equation}
where
\begin{align}
{N_3} =& {a_o}j\omega({(j\omega)^{2\gamma }} + 3{\omega _o}{(j\omega)^\gamma } + 3{\omega _o}^2) \nonumber\\
{D_3} = &{a_o}j\omega({(j\omega)^{2\gamma }} + 3{\omega _o}{(j\omega)^\gamma } + 3{\omega _o}^2)+ {(j\omega)^\gamma }({(j\omega)^{2 + \gamma }} \nonumber\\
&+ 3{\omega _o}{(j\omega)^2} + 3{\omega _o}^2{(j\omega)^\gamma } + {\omega _o}^3)
\label{eq_65}
\end{align}
From (\ref{eq_43}) and (\ref{eq_60}), the mean-square error between the  fractional-order integrator and $P_{fo}$ can be expressed as follows:
\begin{equation}
{\Delta_{fo}} = \frac{{{N_4}}}{{{D_4}}}
\label{eq_66}
\end{equation}
where
\begin{align}
{N_4} =& {\omega _o}^3{(j\omega)^\gamma } - {(j\omega)^\gamma }{({(j\omega)^\gamma } + {\omega _o})^3} \nonumber\\
& + ({a_o}j\omega+ {(j\omega)^2})({(j\omega)^{2\gamma }} + 3{\omega _o}{(j\omega)^\gamma } + 3{\omega _o}^2)\nonumber\\
{D_4} =& {\omega _o}^3{(j\omega)^\gamma } + ({a_o}j\omega \nonumber\\
& + {(j\omega)^2})({(j\omega)^{2\gamma }} + 3{\omega _o}{(j\omega)^\gamma } + 3{\omega _o}^2)
\label{eq_66_}
\end{align}

\begin{figure}[!ht]
	\centering
	\includegraphics[scale=0.50]{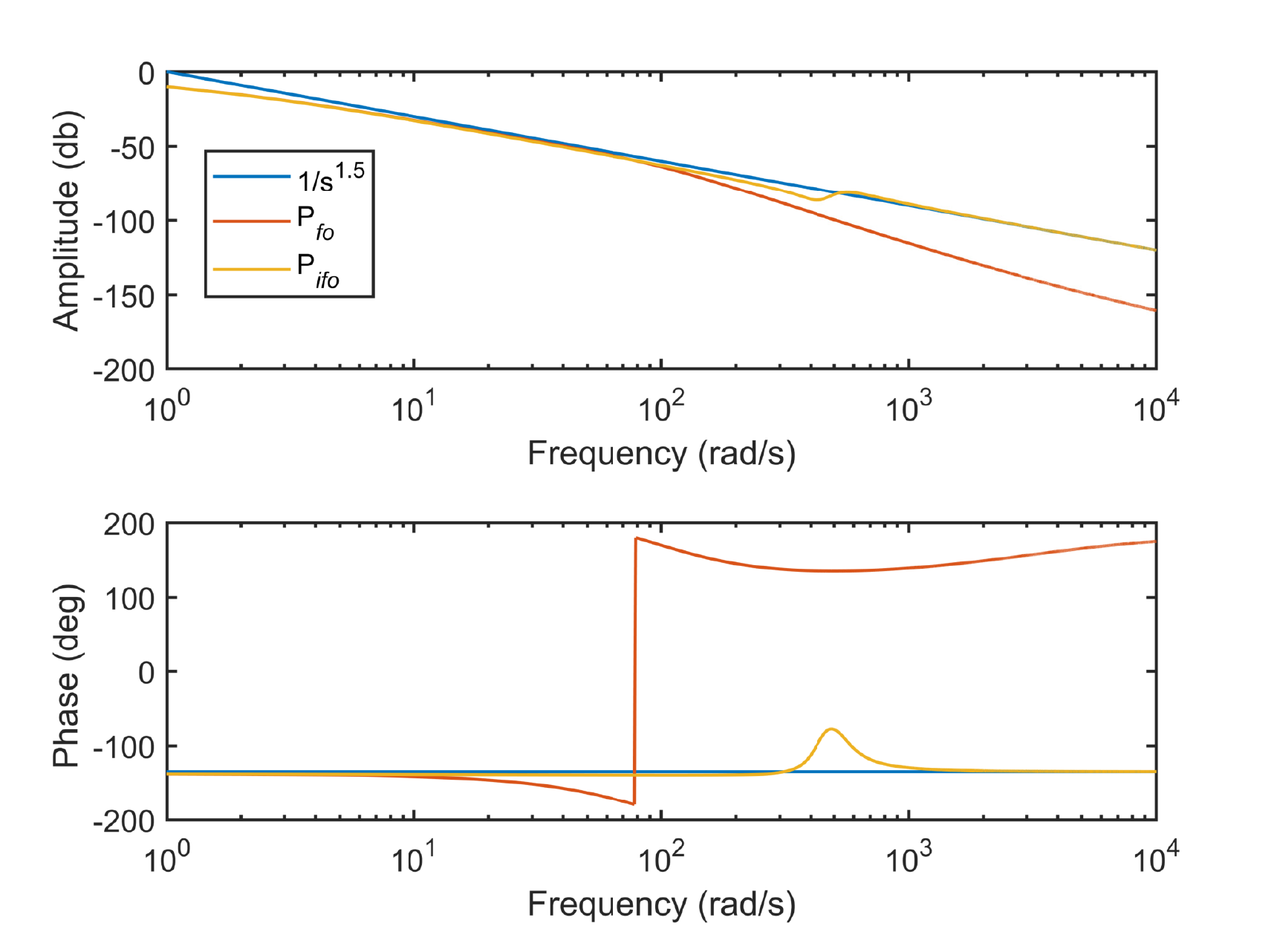}
	\caption{Bode diagram of the  model $Y(s)/U_o(s)$:   $P_{ifo}$  in (\ref{eq_57_}) with IFO-ESO;  $P_{fo}$  in (\ref{eq_60}) with  FO-ESO.  }
	\label{figure_4_1}
\end{figure}

\begin{figure}[t]
	\centering
	\includegraphics[scale=0.55]{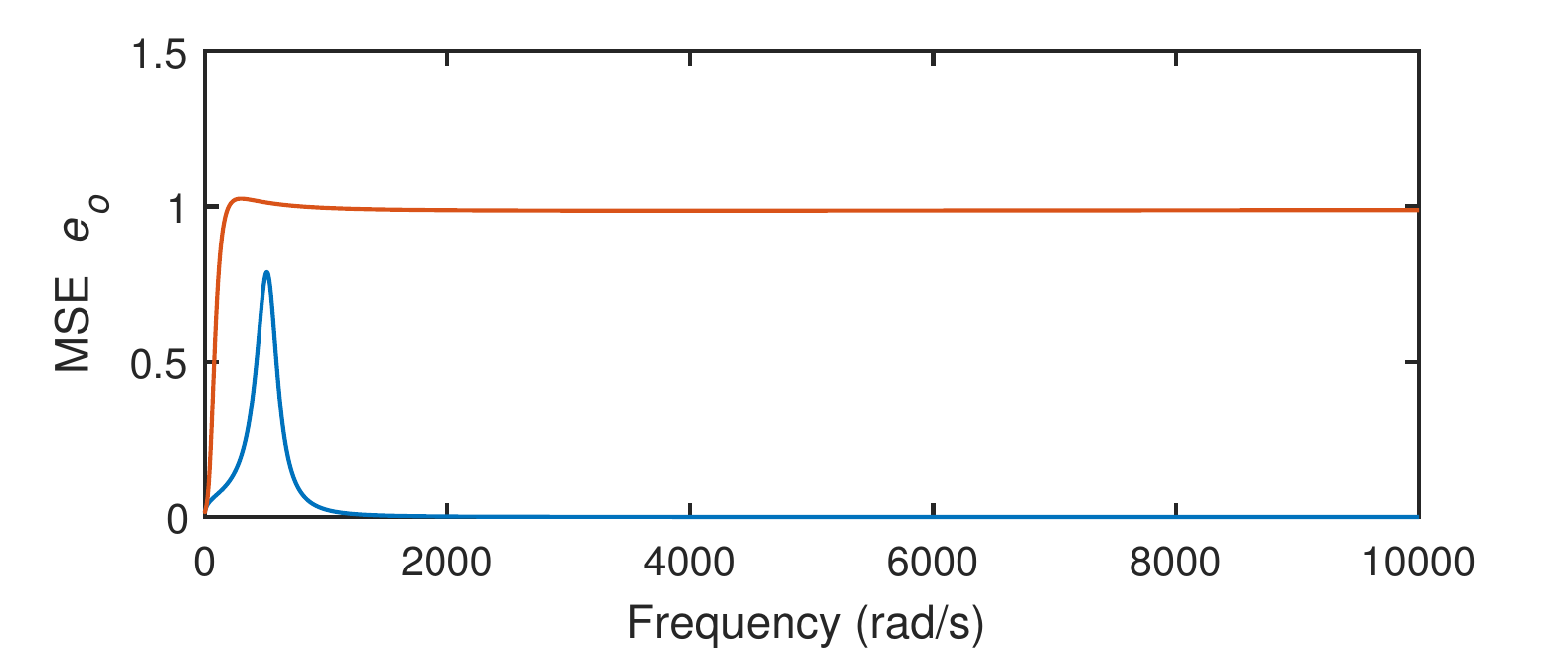}
	\caption{The MSE curves when $a_o = 26.08$, $\gamma = 0.75$, and $\omega_o = 700$}
	\label{figure_5_4}
\end{figure}

Now, let us use the Bode diagram to study the performance of FO-ESO and IFO-ESO with observer gains $L=[3\omega_0,3\omega_0^2,\omega_0^3]\t$. Moreover, the system parameters and the observer gain parameters are adapted from  \cite{chen2021fractional}, i.e.,
$a_o = 26.08$, $b_0=383.635$,  $\omega_o = 700$ and $\gamma = 0.75$. The Bode diagram of $P_{ifo}$ in (\ref{eq_57_}) and $P_{fo}$  in (\ref{eq_60}) are illustrated in Fig.~\ref{figure_4_1}.  
It is clearly shown in Fig.~\ref{figure_4_1} that the amplitude and the phase diagram of  $P_{ifo}$ are close   to  that of $1/s^{1.5}$ within  a larger fequency band  than that of $P_{fo}$ are. In particular, the approximation of $P_{fo}$ to $1/s^{1.5}$ gets worse in high-frequency band. Fig.~\ref{figure_5_4} shows  the mean-square error $e_{fo}$ and $e_{ifo}$ with respect to frequency $\omega$. The result coincides with Fig.~\ref{figure_4_1} that $P_{ifo}$ better approximates the integrator $1/s^{1.5}$ than $P_{fo}$ does. In other words, the IFO-ESO has better performance in terms  of disturbance estimation than the FO-ESO. Therefore, the closed-loop system resulting from the IFO-ESO is more robust than that from the FO-ESO.

\begin{figure}[!ht]
	\centering
	\includegraphics[scale=0.55]{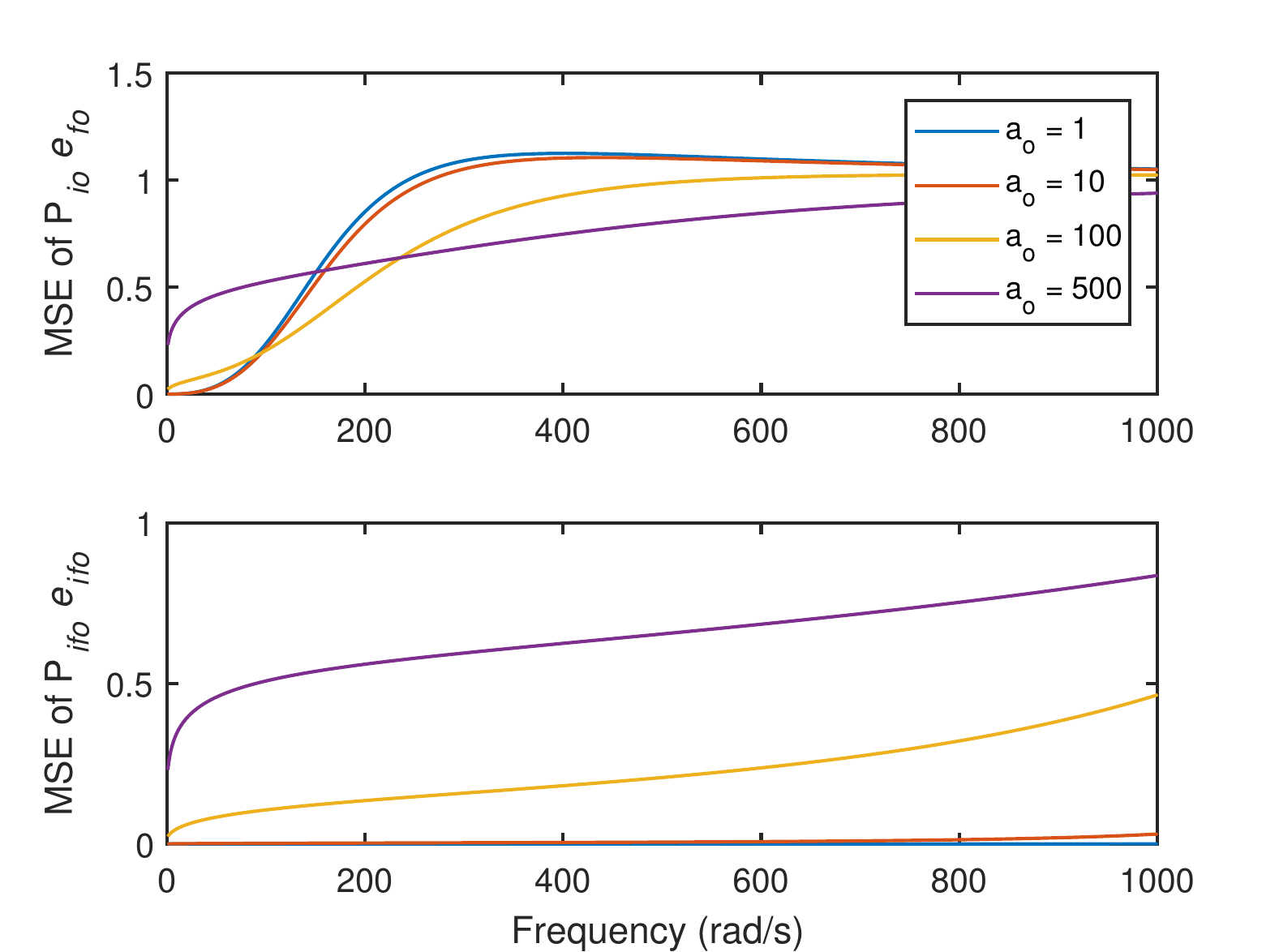}
	\caption{The MSE curves with different $a_o$ when $\omega_o$ = 2000 and $\gamma$ = 0.75}
	\label{figure_5_2}
\end{figure}

\begin{figure}[!ht]
	\centering
	\includegraphics[scale=0.55]{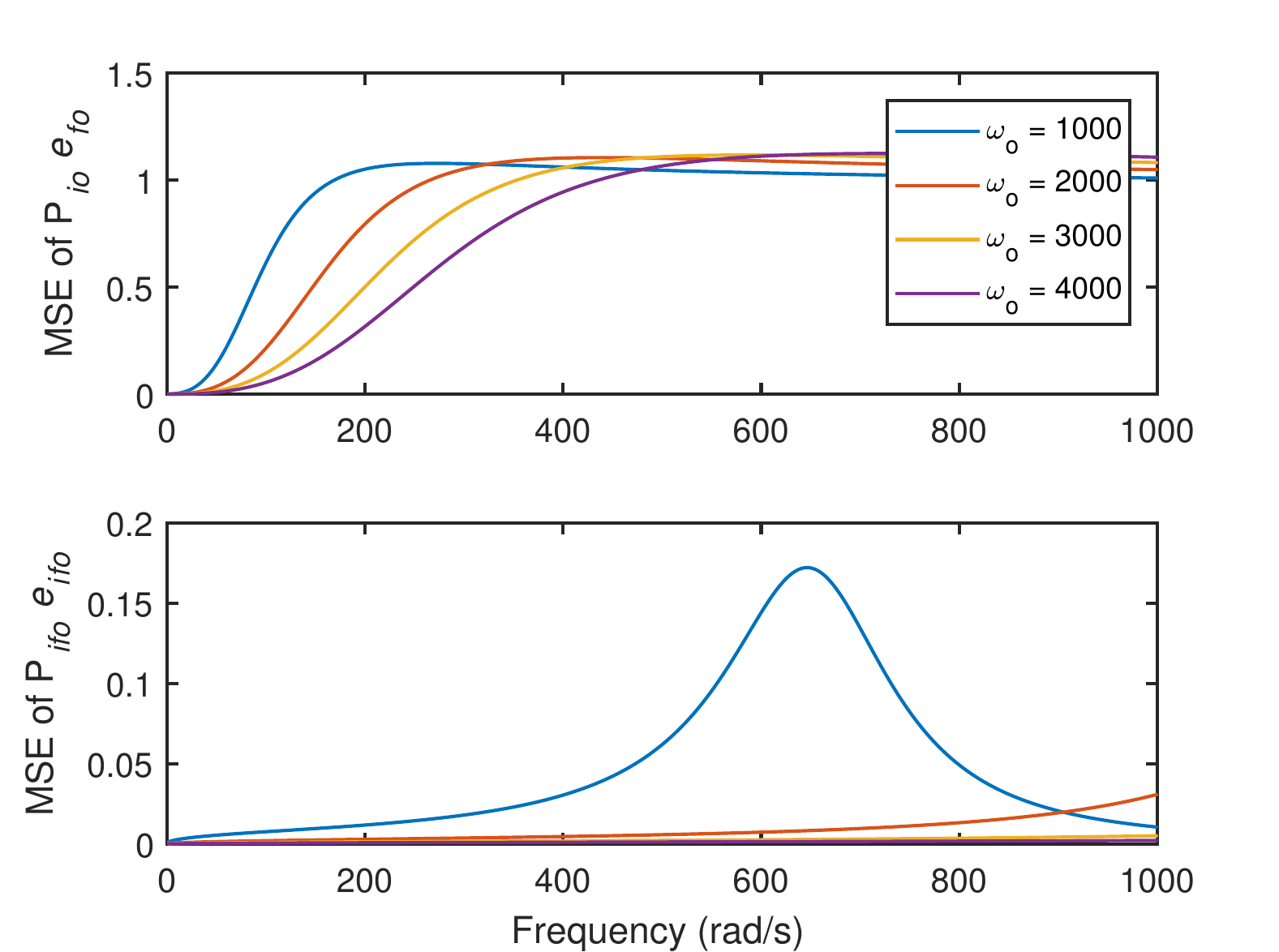}
	\caption{The MSE curves with different $\omega_o$ when $a_o$ = 10 and $\gamma$ = 0.75}
	\label{figure_5_3}
\end{figure}

\begin{figure}[!ht]
	\centering
	\includegraphics[scale=0.55]{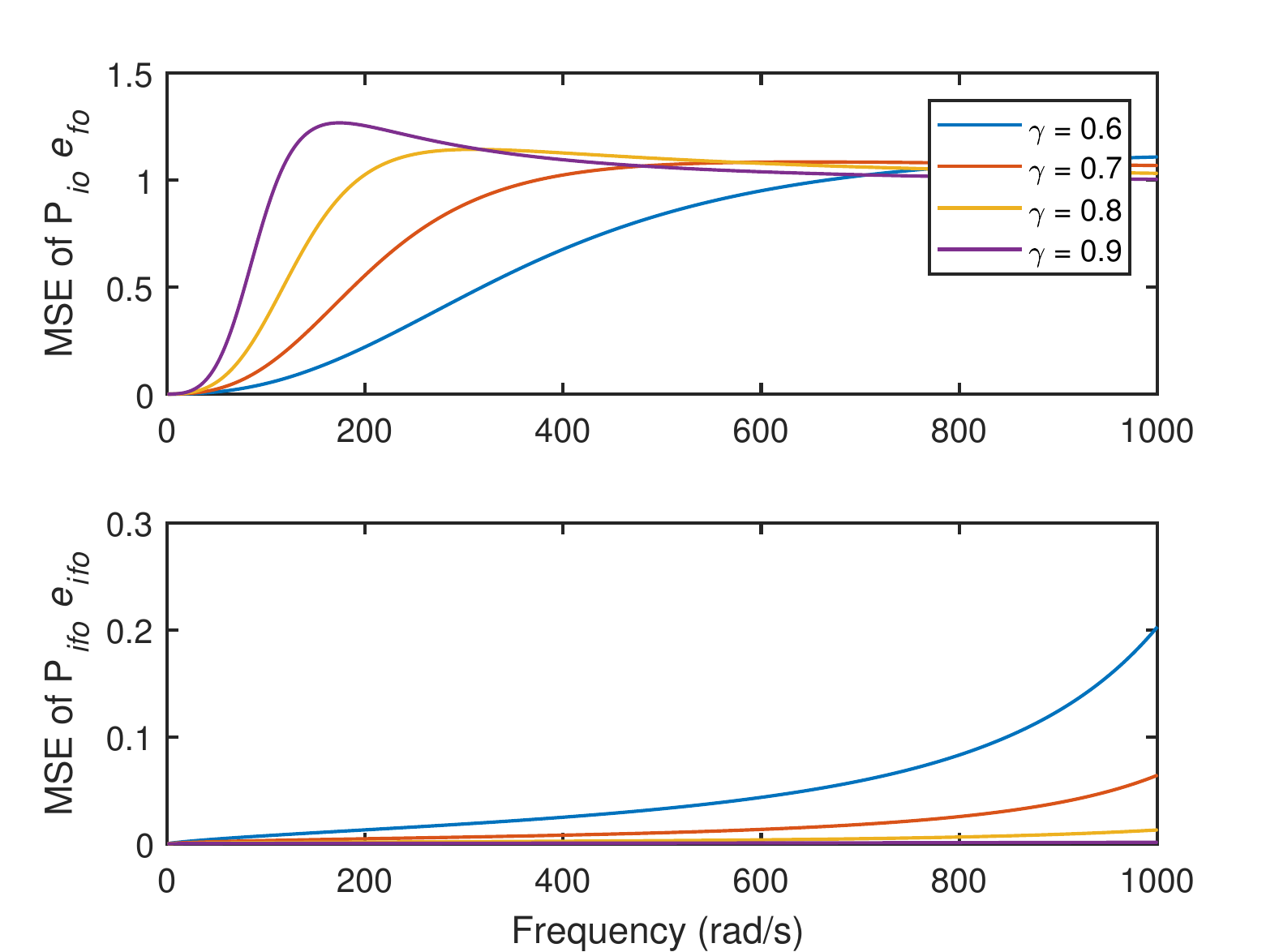}
	\caption{The MSE curves with different $\gamma$ when $a_o$ = 10 and $\omega_o$ = 2000}
	\label{figure_5_5}
\end{figure}

Fig.~\ref{figure_5_2} shows the curves of the mean-square error $e_{fo}$ and $e_{ifo}$ with different model paramter $a_o$ when $\omega_o= 2000$  and $\gamma= 0.75$. Fig.~\ref{figure_5_3} shows the curves of the mean-square error $e_{fo}$ and $e_{ifo}$ with different observer gains $\omega_o$ when $a_o= 10$  and $\gamma= 0.75$.  Fig.~\ref{figure_5_5} demonstrates the variation of the mean-square error with different order $\gamma$ when $a_o= 10$  and $\omega_o= 2000$. As shown in Fig.~\ref{figure_5_2}, Fig.~\ref{figure_5_3}, and Fig.~\ref{figure_5_5}, the mean-square error $e_{ifo}$  is less prone to the variation of system parameter $a_o$, controller parameters $\omega_o$, and the order $\gamma$ than  the mean-square error $e_{fo}$ is.

\brem
Note that if the performance of ESO is  robust to the variation of the closed-loop system parameters $a_o$,  $\omega_o$, and $\gamma$, the auxiliary tracking controller  design of $u_0$ based on the compensated system (\ref{eq_14}) or (\ref{eq:FO-App})
can achieve better performance when these parameters vary. In other words, the IFO-ADRC closed-loop system is more robust with respect to the variations of the plant $G(s)$ parameters and ESO parameters.\erem

\section{Time-domain simulation and Comparison \label{sec:TDS}}
\begin{figure}[!ht]
	\centering
	\includegraphics[scale=0.042]{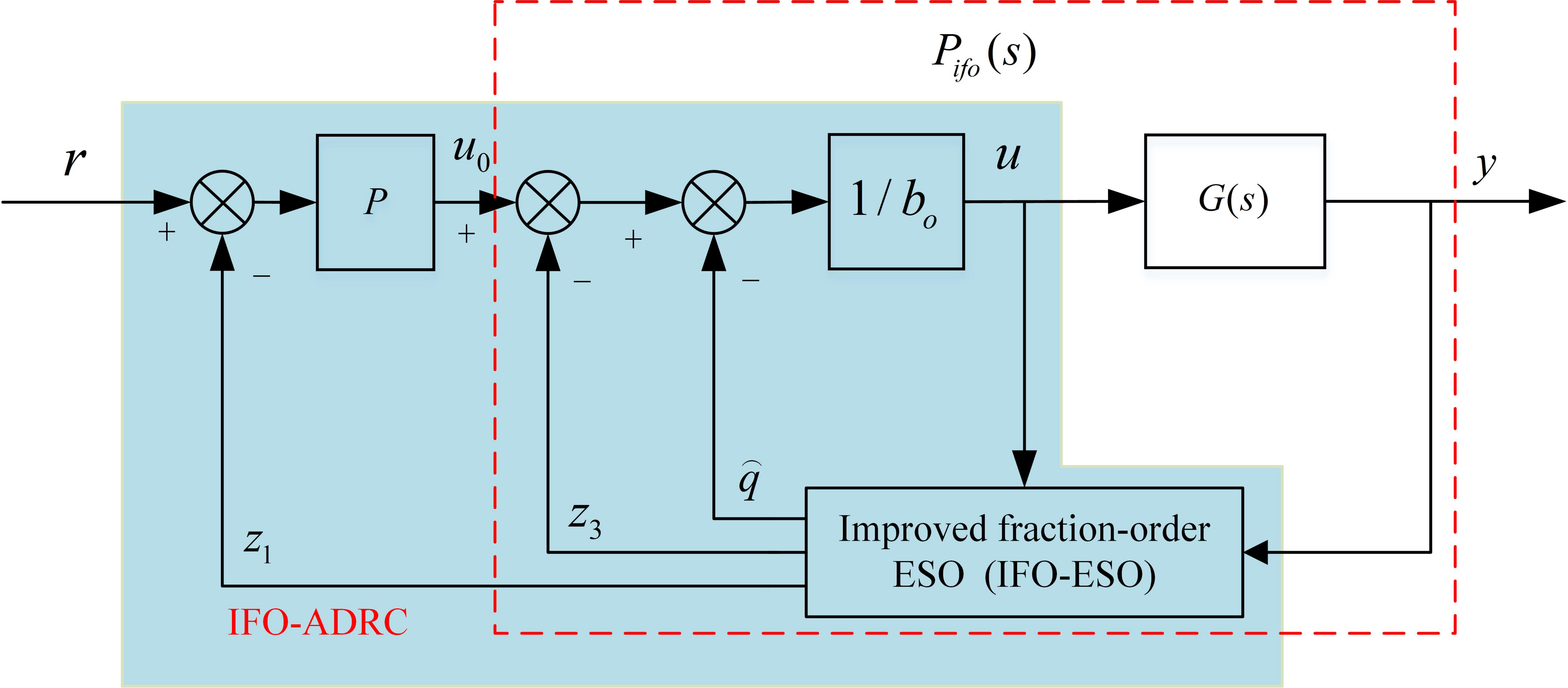}
	\caption{Structure of the IFO-ADRC}
	\label{figure_3_1}
\end{figure}

\begin{figure}[!ht]
	\centering
	\includegraphics[scale=0.042]{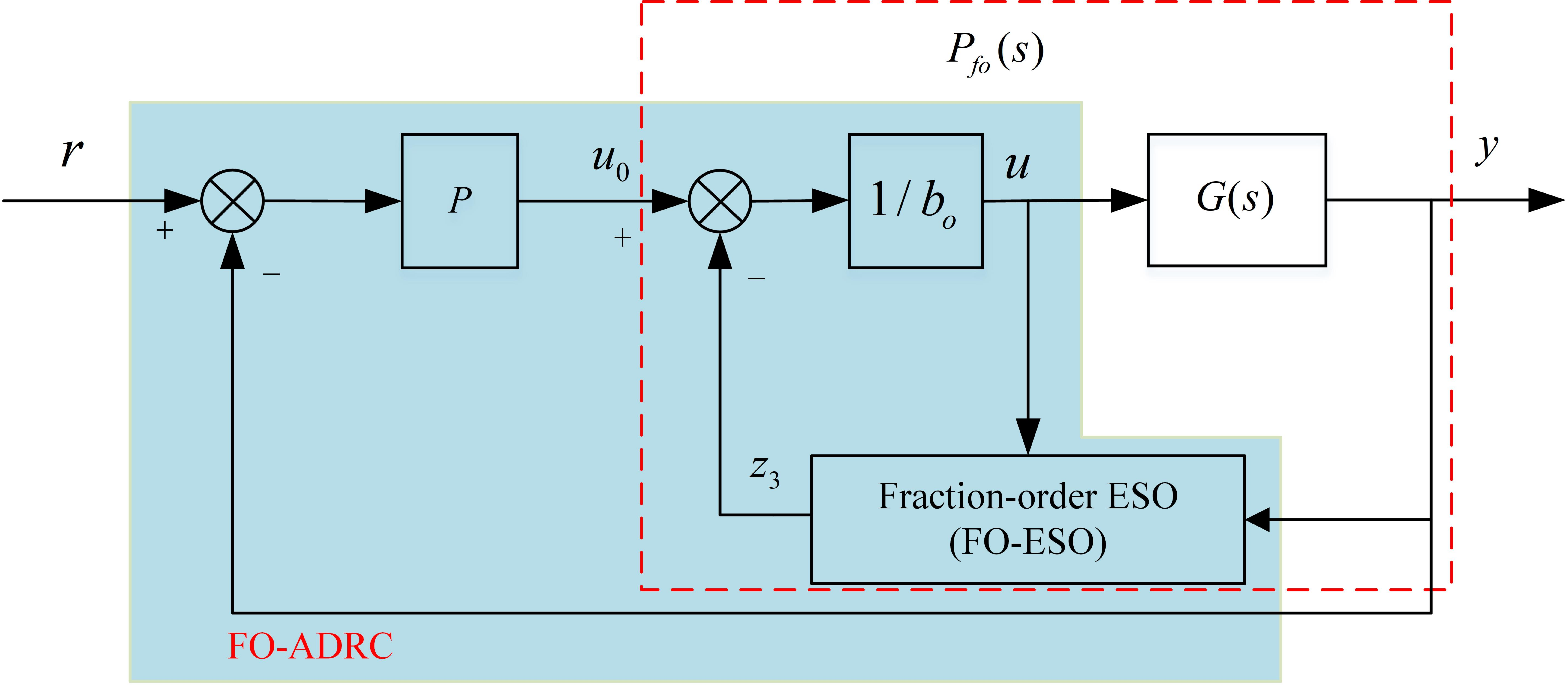}
	\caption{Structure of the FO-ADRC in \cite{chen2021fractional}}
	\label{figure_2_1}
\end{figure}

\begin{figure}[!ht]
	\centering
	\includegraphics[scale=0.042]{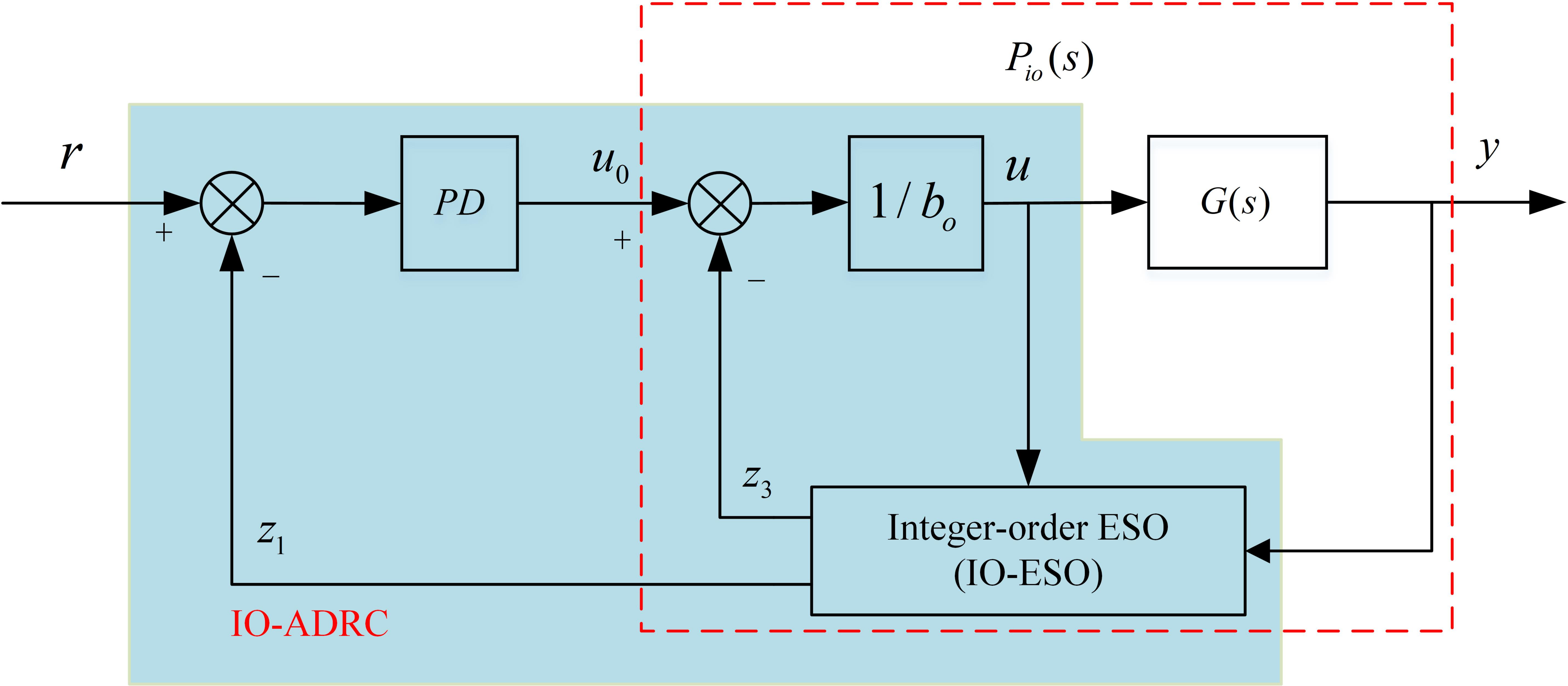}
	\caption{Structure of the IO-ADRC}
	\label{figure_3_2}
\end{figure}

In this section, we will show  the performance  of the IFO-ADRC in the time-domain using MATLAB/Simulink and compare it with IO-ADRC and FO-ADRC. The plant  used for the simulation is  (\ref{eq_53_})  which is adopted from the example of \cite{chen2021fractional}.
The structures of IFO-ADRC, FO-ADRC and  IO-ADRC are similar and presented in Fig.~\ref{figure_3_1}, \ref{figure_2_1} and \ref{figure_3_2}, respectively. 
An advantage of IFO-ADRC and FO-ADRC  is that a simpler  auxiliary controller $u_0$ can be used.
Note that the auxiliary controller $u_0$ for IFO-ADRC and FO-ADRC  is a P controller,  while PD controllers must be used in IO-ADRC to ensure the stability.   
The  controller in IFO-ADRC and FO-ADRC is
	\begin{equation} \label{eq:CP}
	{C_{p}}(s) = {K_{fp}},
	\end{equation}
	where
	$K_{fp}$ is the parameter of P controller
	and the PD controller in IO-ADRC is
	\begin{equation}
	{C_{pd}}(s) = {K_{ip}}( 1 + {K_{id}}s),
	\end{equation}
	where $K_{ip}$ and $K_{id}$ are the PD  parameters.  The IO-ESO used in IO-ADRC is given \cite{gao2006scaling},

	\begin{gather}
	\dot{z}=Az+Bu+L(y-\upfrown{y})\nonumber\\
	\upfrown{y}=Cz \label{eq:IO-ESO2}
	\end{gather}
	Refer to (\ref{eq:FO-ESO3}) for the matrices $z$, $A$, $B$, and $C$. The controller $u$ in IO-ADRC is given 
	\begin{equation} 
	u = \frac{u_0  - \upfrown{f}_{io}}{b_0}
	\end{equation}

The observer gains $L=[\beta_1,\beta_2,\beta_3]\t=[3\omega_0,3\omega_0^2,\omega_0^3]\t$ and the order $\gamma$ are adopted from the reference paper \cite{chen2021fractional}, i.e.,   $a_o = 26.08$, $b_0 = 383.635$, $b=b_0$, ${\omega _o} = 700 $ rad/s, and $\gamma=0.75$. The fractional-order  operator ${s^\gamma }$ is discretized by the impulse response invariant method \cite{mis} where the discrete frequency for IFO-ESO and FO-ESO is 8000 Hz and the discrete order of the fractional-order operators is 7. Let the P controller parameter be $K_{fp} = 356$.  Note that 
the observer  parameters satisfy  conditions of Theorem \ref{th_closed-loop}  and Theorem 1 in \cite{chen2021fractional}, the IFO-ADRC and FO-ADRC closed-loop systems are BIBO stable.

For the fair comparison, based on the open-loop transfer function, the PD controller used in IO-ADRC parameters are chosen to ensure that the gain crossover frequency $\omega_c^*$ and the phase margin ${\varphi _m}$ of the IO-ADRC are the same as that of the IFO-ADRC \cite{chen2021fractional,zheng2021synthesis}. For IFO-ADRC, the open-loop transfer function can be described as
	\begin{equation}
	{G_{ifo}}(s) = C_p(s)\frac{Z_1(s)}{U_0(s)}= K_{fp}{P_{ifo}}(s)\frac{{{Z_1}(s)}}{{Y(s)}}
	\end{equation}
where
\begin{gather}
\frac{{{Z_1}(s)}}{{Y(s)}} = \frac{{{s^{2 + \gamma }} + 3{\omega _o}{s^2} + {a_o}{s^{1 + \gamma }} + 3{\omega _o}^2{s^\gamma } + {\omega _o}^3}}{{{s^{2 + \gamma }} + 3{\omega _o}{s^2} + 3{\omega _o}^2{s^\gamma } + {\omega _o}^3}}
\end{gather}
The gain crossover frequency  $\omega^*_c$ = 42 rad/s and the phase margin ${\varphi _m}={39.9^ \circ}$
can be calculated from
\begin{gather}
\angle |{G_{ifo}}(j{\omega _c}^*)| =  - \pi  + {\varphi _m}\nonumber\\
|{G_{ifo}}(j{\omega _c}^*)| = 1
\end{gather}
For IO-ADRC, the open-loop transfer function is
\begin{equation}
{G_{io}}(s) =  {C_{pd}}(s){P_{io}}(s)\frac{{{Z_1}(s)}}{{Y(s)}}
\end{equation}
where
\begin{gather}
{P_{io}}(s) =\frac{{Y(s)}}{{{U_0}(s)}} = \frac{{{{(s + {\omega _o})}^3}}}{{s({a_o} + s){{(s + {\omega _o})}^3} - {a_o}{\omega _o}^3s}} \nonumber\\
\frac{{{Z_1}(s)}}{{Y(s)}} = \frac{{{s^3} + (3{\omega _o} + {a_o}){s^2} + 3{\omega _o}^2s + {\omega _o}^3}}{{{s^3} + 3{\omega _o}{s^2} + 3{\omega _o}^2s + {\omega _o}^3}}
\end{gather}
Then, the PD controller for  IO-ADRC can be designed as
\begin{equation}
{C_{pd}}(s) = 1559.83(1 + 0.0199s)
\end{equation} 
to ensure that the gain crossover frequency $\omega_c^*=42$ rad/s and the phase margin ${\varphi _m}=39.9^ \circ$

\begin{figure}[t]
	\centering
	\includegraphics[scale=0.55]{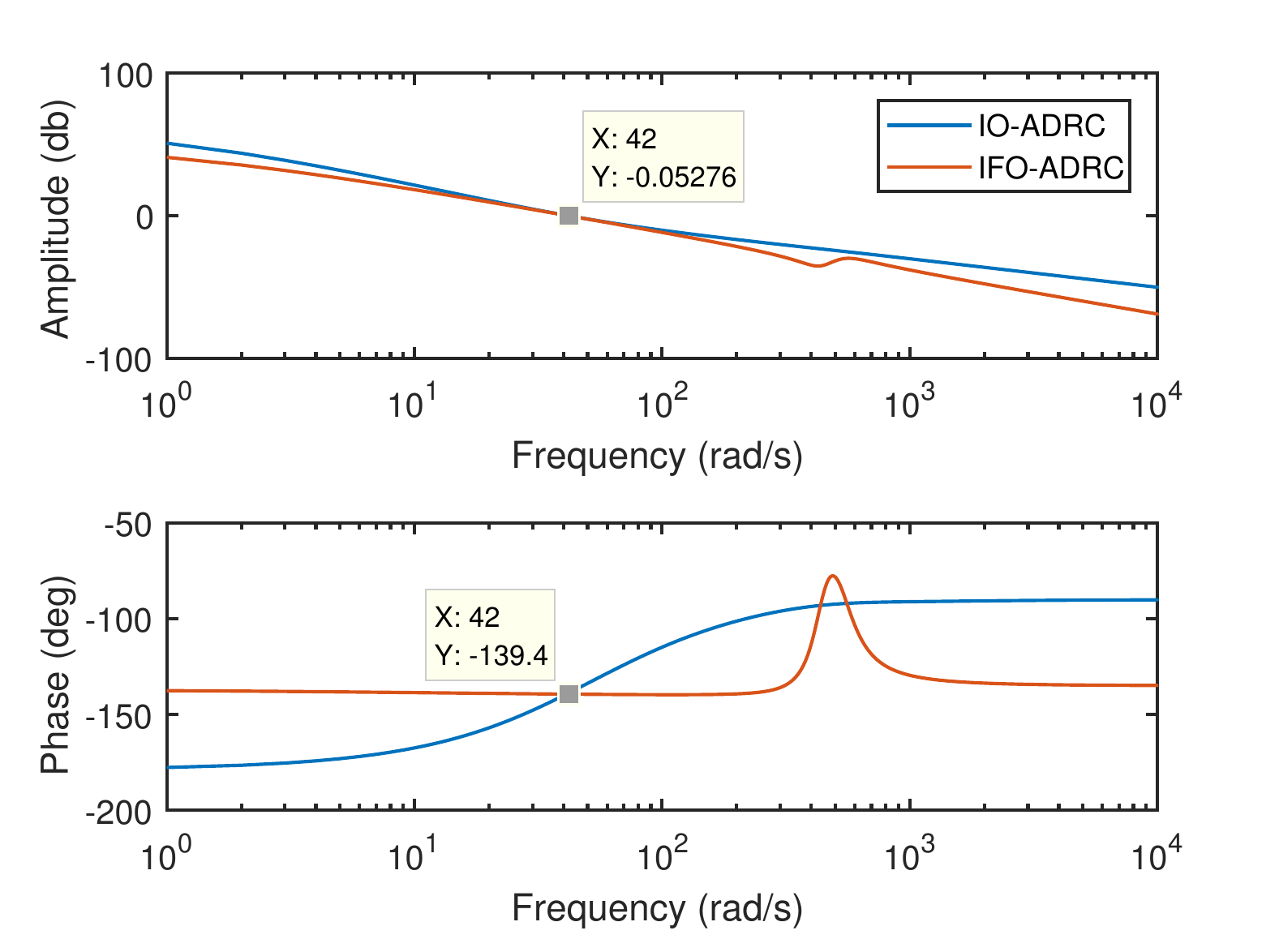}
	\caption{Open-loop Bode diagram of the IO-ADRC system and the IFO-ADRC system}
	\label{figure_1}
\end{figure}

The open-loop Bode diagram  of the IO-ADRC and IFO-ADRC systems is illustrated in Fig.~\ref{figure_1} showing that the two systems have the same  gain crossover frequency $\omega^*_c$ = 42 rad/s and phase margin ${\varphi _m}={39.9^ \circ}$.  From the open-loop Bode diagram of the IO-ADRC system, the IO-ADRC closed-loop system is stable.

\begin{table}[!t]
	\renewcommand{\arraystretch}{1.3}
	\caption{Comparison of the responses with three control systems (simulation)}
	\centering
	\includegraphics[scale=0.080]{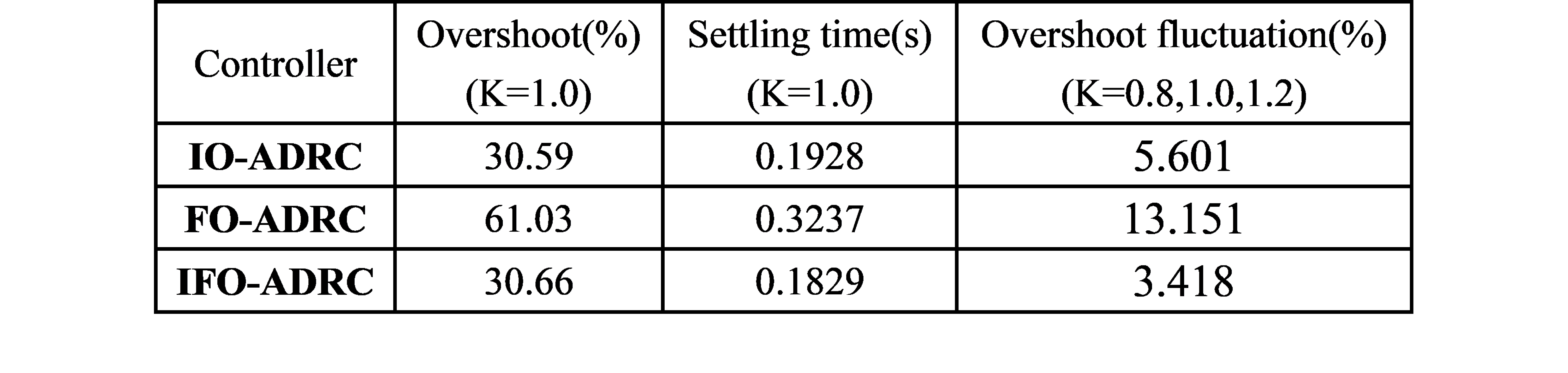}
	\label{table_1}
\end{table}

\begin{figure}[t]
	\centering
	\includegraphics[scale=0.058]{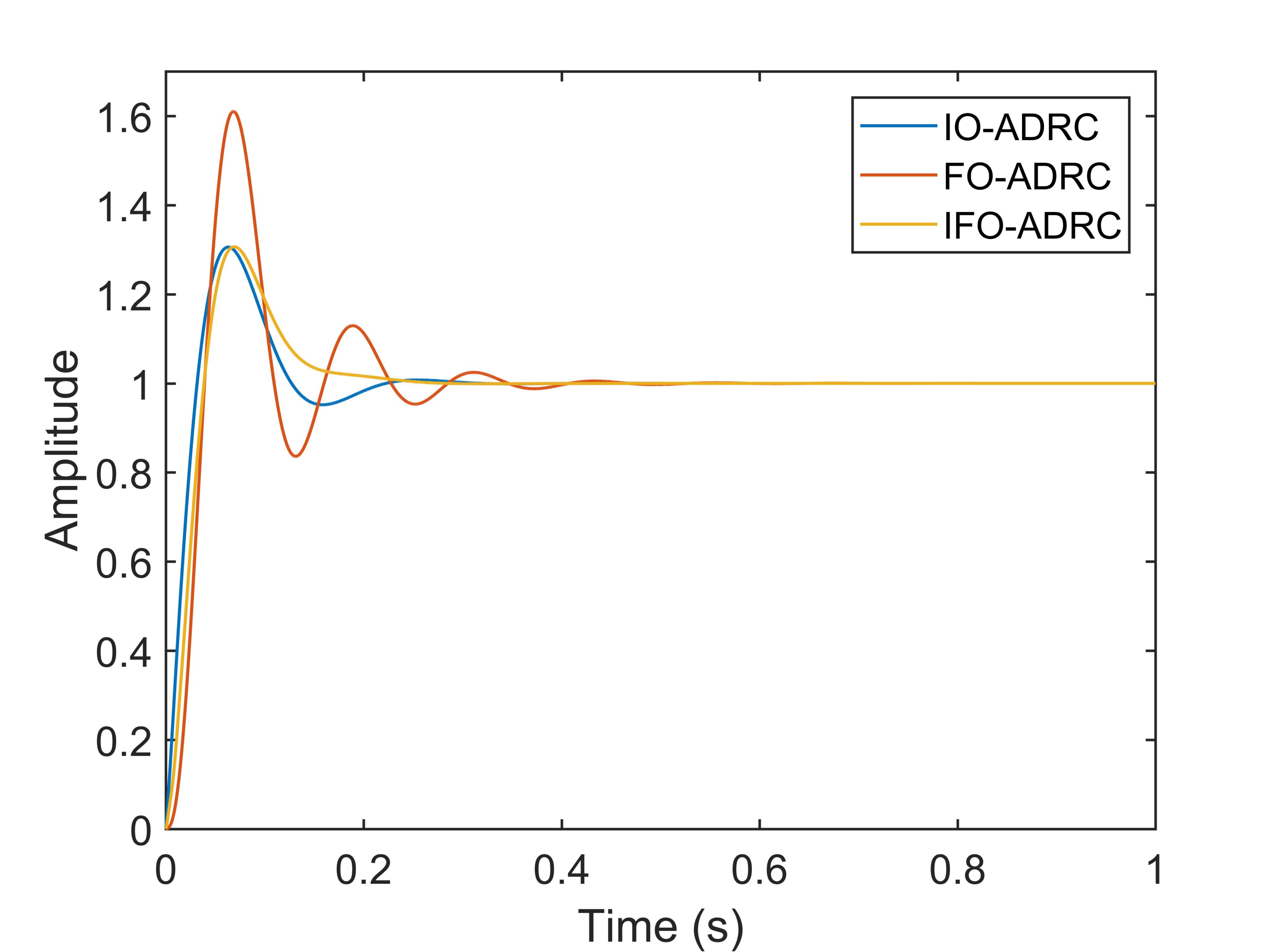}
	\caption{Step responses of three differemt control systems}
	\label{figure_4_2}
\end{figure}

The step responses of the IO-ADRC,  FO-ADRC, and IFO-ADRC systems are shown in Fig.~\ref{figure_4_2}. It is shown in Fig.~\ref{figure_4_2} that the IFO-ADRC   system has better dynamic response performance than the FO-ADRC and   IO-ADRC   systems. The IFO-ADRC   system has smaller overshoot than the FO-ADRC system.

\begin{figure}[t]
	\centering
	\includegraphics[scale=0.058]{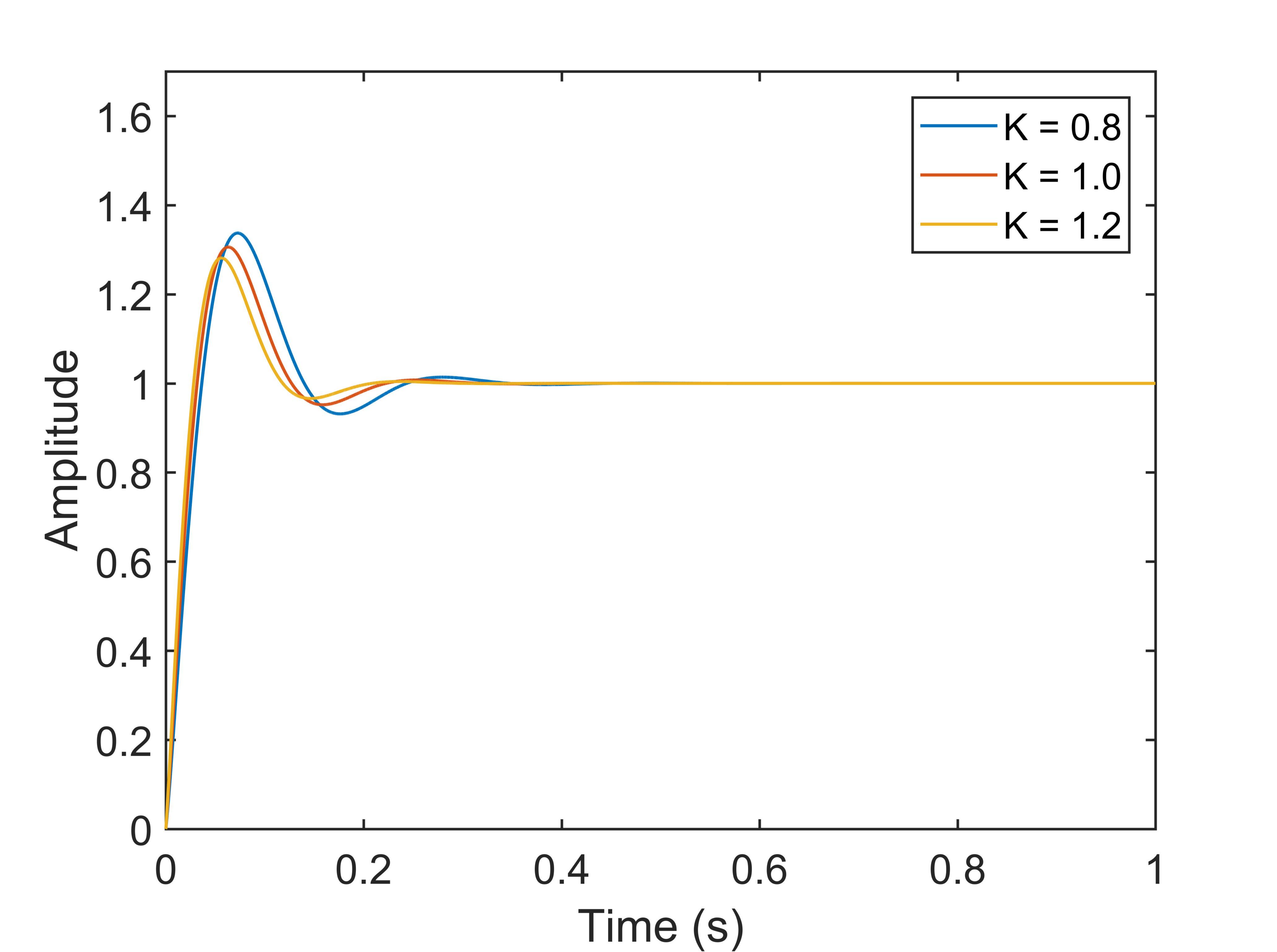}
	\caption{Step responses of the IO-ADRC control system with controller gain variations}
	\label{figure_5_1}
\end{figure}

\begin{figure}[t]
	\centering
	\includegraphics[scale=0.058]{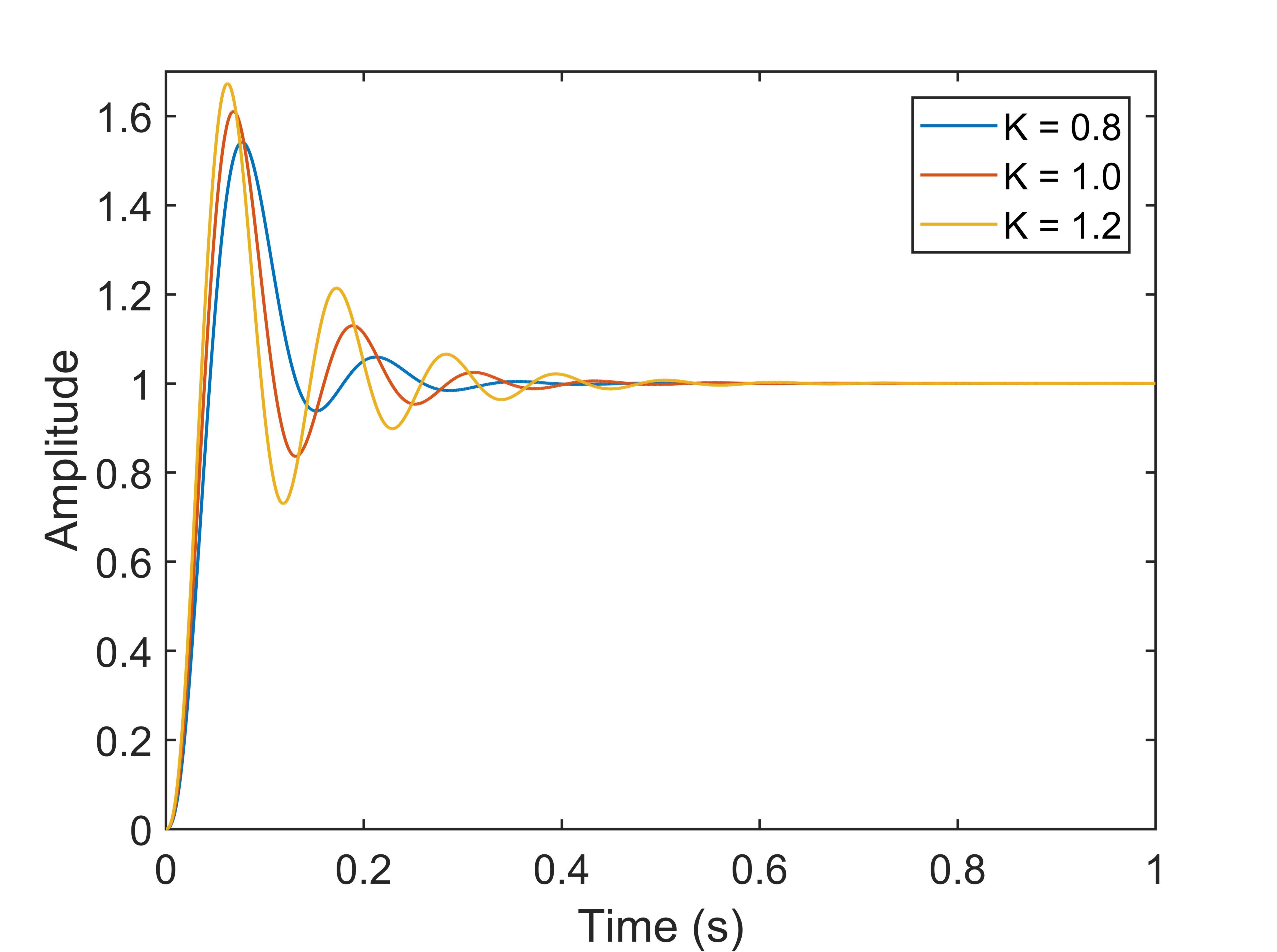}
	\caption{Step responses of the FO-ADRC  control system with controller gain variations}
	\label{figure_6_1}
\end{figure}

\begin{figure}[t]
	\centering
	\includegraphics[scale=0.058]{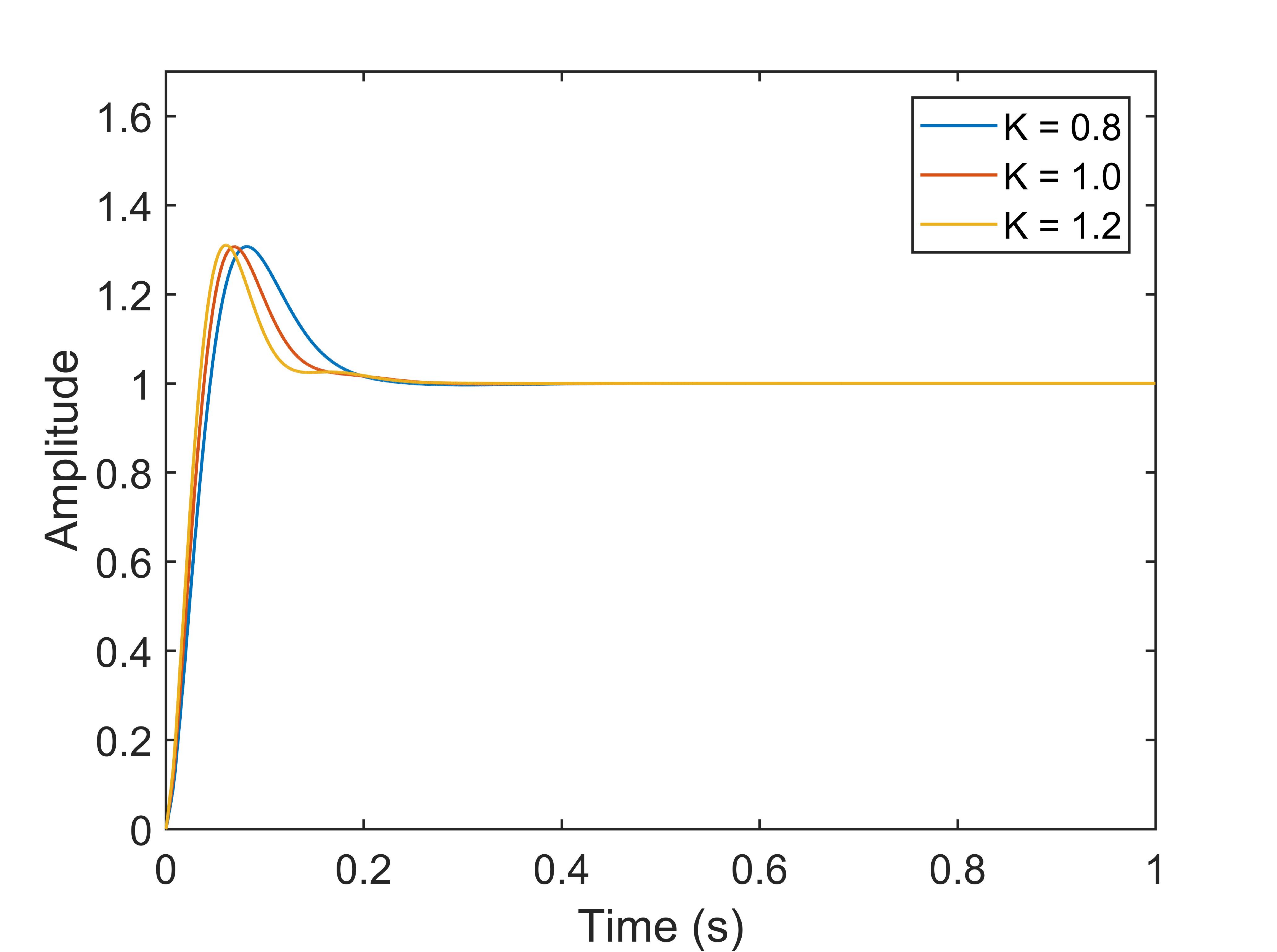}
	\caption{Step responses of the IFO-ADRC control system with controller gain variations}
	\label{figure_7_1}
\end{figure}

Now, let us consider the system performance variation against 
controller parameters. Let us multiple the P controller parameter $K_{fp}$ and PD controller parameter 
$K_{ip}$ by $K=0.8$ and $K=1.2$. The nominal value is used when $K=1$.
Fig.~\ref{figure_5_1}, Fig.~\ref{figure_6_1} and Fig.~\ref{figure_7_1} are the step responses of three closed-loop systems when different controller  parameters  are imposed. As shown in Fig.~\ref{figure_6_1} and Fig.~\ref{figure_7_1}, the IFO-ADRC  system are robust to controller gain variations.

When $K=0.8$, $K=1.0$, or $K=1.2$ are set respectively,  the maximum speed of the step response are denoted as $M_K$. The overshoot fluctuation is calculated as
\begin{gather}
\mbox{overshoot{\;\;}fluctuation{\;  =  \nonumber\;}}\\\frac{{\max \{ {M_{0.8}},{M_{1.0}},{M_{1.2}}\}  - \min \{ {M_{0.8}},{M_{1.0}},{M_{1.2}}\} }}{\mbox{reference{\;\;} input}}
\end{gather}
The step responses of three  closed-loop systems for different $K$ are given in TABLE \ref{table_1}.  Note that the overshoots of IFO-ADRC system and the IO-ADRC system are similar and smaller  than that of the FO-ADRC  system. The settling time of the IFO-ADRC system is shorter than the IO-ADRC and FO-ADRC systems. 


\section{Experiments: PMSM speed servo control}
\begin{figure}[!ht]
	\centering
	\includegraphics[scale=0.045]{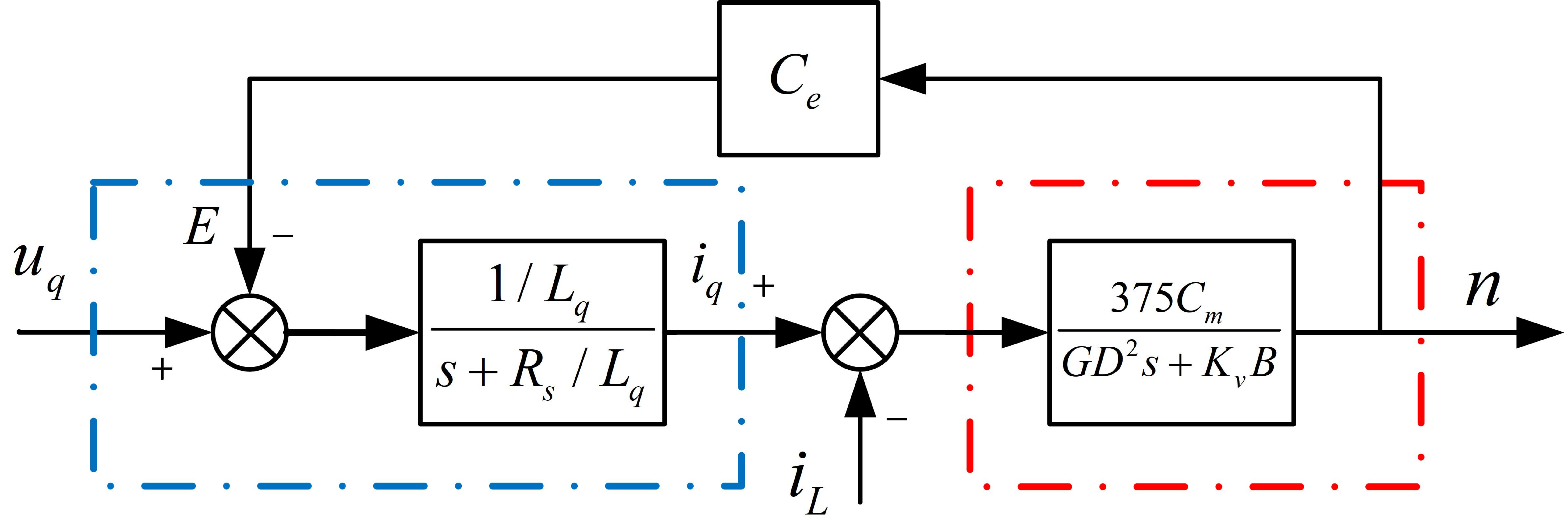}
	\caption{Mathematical model of the PMSM}
	\label{figure_20}
\end{figure}

In the section, the control performance of IFO-ADRC, FO-ADRC, and IO-ADRC  are compared in a real-world application. Experiments are carried out on the PMSM speed servo control system. Fig.~\ref{figure_20} is the block diagram of the PMSM given in $dq$-axis frame. The diagram  encircled by
the blue and red dotted lines are the electromagnetic   and the mechanical part of the PMSM, respectively.

The armature current of DC motor can be represented by  $q$-axis component of the current $i_q$ \cite{zheng2016fractional}. The $q$-axis voltage equation of PMSM is described as follows
\begin{equation}
{u_q} - E = {u_q} - {C_e}n = {R_s}{i_q} + {L_q}\frac{{d{i_q}}}{{dt}}
\label{eq_100}
\end{equation}
where  $u_q$ is the $q$-axis voltage, $E$ is the back
electromotive force, $C_e$ is the electromotive force coefficient, $n$ is the actual speed of motor rotor, $R_s$ is motor phase armature resistance, $i_q$ is $q$-axis current and $L_q$ is $q$-axis inductance.
Taking $u_q$ as the input voltage and $i_q$ as the output current,  the transfer functions of the electromagnetic part  is
\begin{equation}
{G_e}(s) = \frac{{{I_q}(s)}}{{{U_q}(s)}} = \frac{{1/{L_q}}}{{s + {R_s}/{L_q}}}
\end{equation}
where $I_q(s)$ and $U_q(s)$ are the Laplace transforms of $i_q$ and $u_q$, respectively.
Note that  $E$ is regarded as a constant disturbance input of the $q$-axis current loop and is therefore not shown in the transfer function above \cite{ruan2010control}.

The motion equation of PMSM can be described as follows
\begin{equation}
T - {T_L} = {C_m}({i_q} - {i_L}) = \frac{{G{D^2}}}{{375}}\frac{{dn}}{{dt}} + {K_v}Bn\\
\label{eq_101}
\end{equation}
where $T$ is the electromagnetic moment, $T_L$ is the load moment, $C_m$ is the torque coefficient, $i_L$ is the equivalent current of load torque, $G{D^2}$ is flywheel inertia, $K_v$ is a speed conversion factor, $B$ is coefficient of viscous friction, and $n$ is the actual speed of the motor rotor. 
Taking $i_q$ as input current and $n$ as output speed,   the transfer functions of the mechanical part is
\begin{equation}
{G_m} = \frac{{N(s)}}{{{I_q}(s)}} = \frac{{375{C_m}}}{{G{D^2}s + {K_v}B}}
\end{equation}
where $N(s)$ is the Laplace transforms of $n$. 
Note that  the load moment is not considered in this paper, i.e., $i_L = 0$, 

\begin{figure}[htbp]
	\centering
	\subfigure[{Current loop of the PMSM speed servo system}]{
		\includegraphics[scale=0.035]{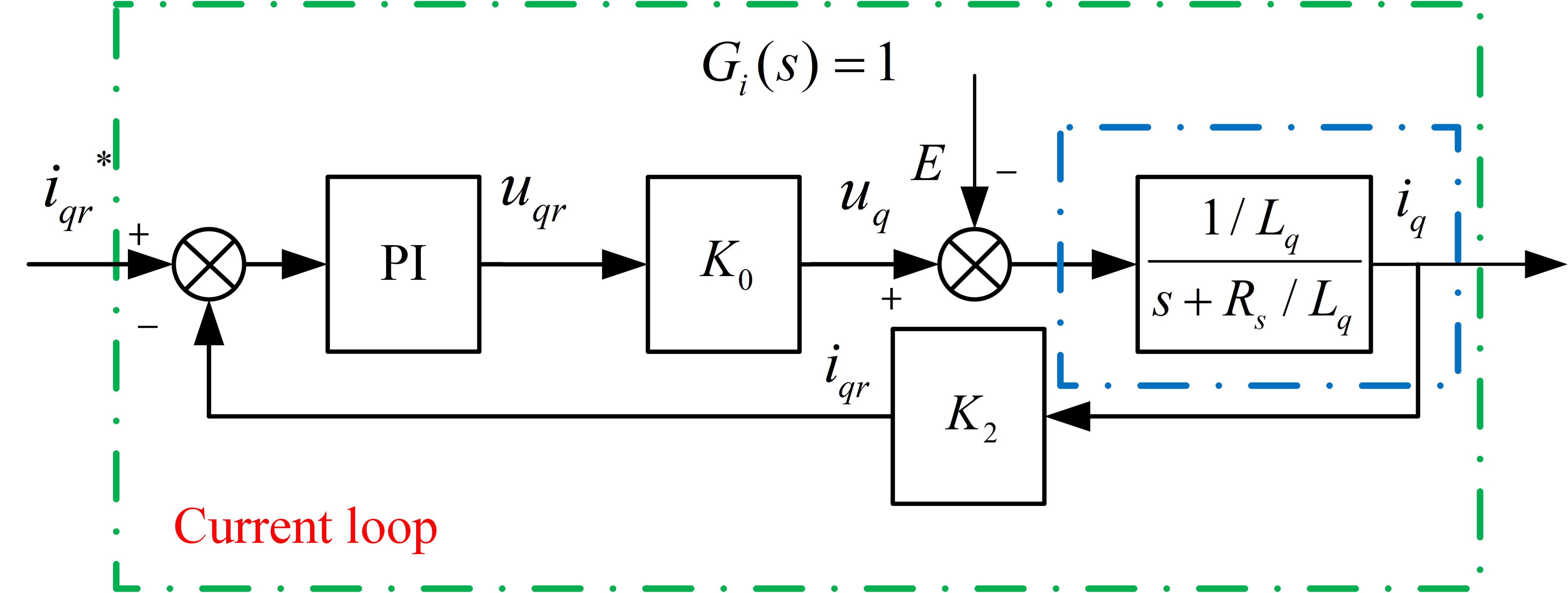} \label{1}
	}
	\quad
	\subfigure[{ Speed loop of the PMSM speed servo system}]{
		\includegraphics[scale=0.035]{speed_loop.eps} \label{2} 
	}
	\caption{PMSM speed servo system using IFO-ADRC}
	\label{figure_21}
\end{figure}

The block encircled by the green dash-dotted line in Fig.~\ref{figure_21}(a) is the block diagram of the current loop $G_i(s)$ of the PMSM speed servo system.  In Fig.~\ref{figure_21}(a), $u_{qr}$ and $i_{qr}$ are the per unit of the actual voltage and the actual current, respectively, $i_{qr}^*$ is the reference input of the current loop, $K_0$ and $K_2$ are the voltage and current conversion factors, respectively. The PI controller in current loop is designed to ensure that $G_i(s) = 1$ in the operating frequency band of the speed loop. Fig. \ref{figure_21}(b) is the block diagram of the speed loop of the PMSM speed servo system using IFO-ADRC. In Fig. \ref{figure_21}(b), $K_1$ is the speed conversion factor, $T_i$ is the speed feedback filter coefficient, and $n_r$ is the per unit of the actual speed.
Since $G_i(s) = 1$, thus the plant of the speed loop   is
\begin{equation}
G(s) =  \frac{{{N_r}(s)}}{{{I_q}(s)}} = \frac{{375{C_m}{K_1}}}{{{T_i}G{D^2}{s^2} + ({K_v}B{T_i} + G{D^2})s + {K_v}B}}
\label{eq_45}
\end{equation}   
where $N_r(s)$ and $I_q(s)$ are Laplace transforms of $n_r$ and $i_q$, respectively. 

\begin{figure}[t]
	\centering
	\includegraphics[scale=1.0]{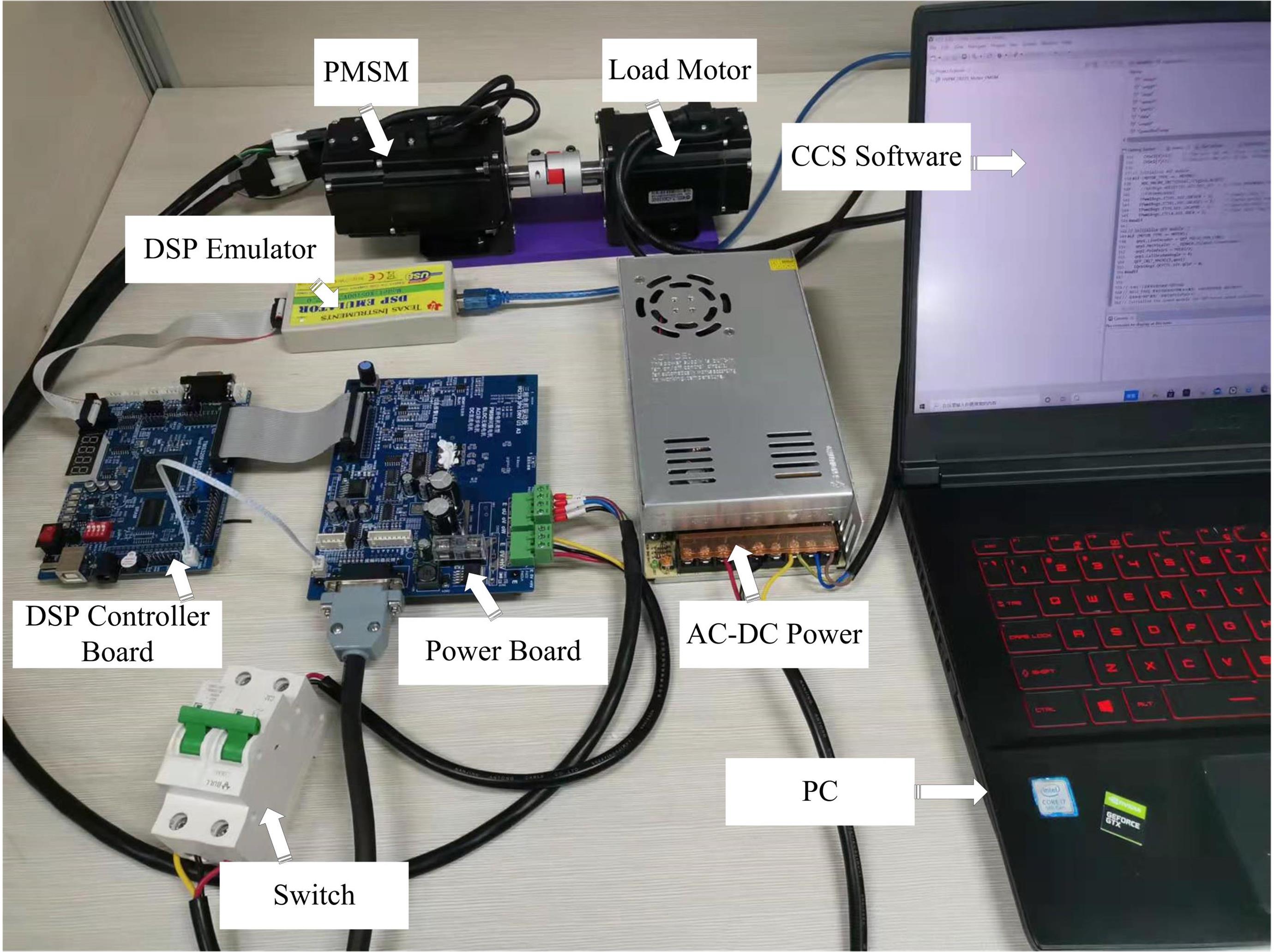}
	\caption{Experimental platform for control performance validation}
	\label{figure_9}
\end{figure}

The closed-loop controller is implemented on digital signal processor (DSP)  illustrated in Fig.~\ref{figure_9}. The PMSM is 60ST-M00630C and MOSFET is adopted as the gate driver.  The specification of the PMSM is shown in TABLE \ref{table_3}. $K_0 = 20.3$ is determined by the actual hardware, $K_1 = 1/1200$, $K_2 = 1/9.9$, $T_i = 1/100$ are configured by software, and $K_v = 30/\pi$ is the speed conversion factor from rad/s to r/min. The speed sampling period was set as $0.125$ ms, and the current loop sampling period is set as $0.0625$ ms. The motor speed waveform is collected by DSP Emulator and CCS software.
{ \blue
	
	\begin{table}[!ht]
		\renewcommand{\arraystretch}{1.3}
		\caption{   The specification of the PMSM }
		\centering
		\includegraphics[scale=0.085]{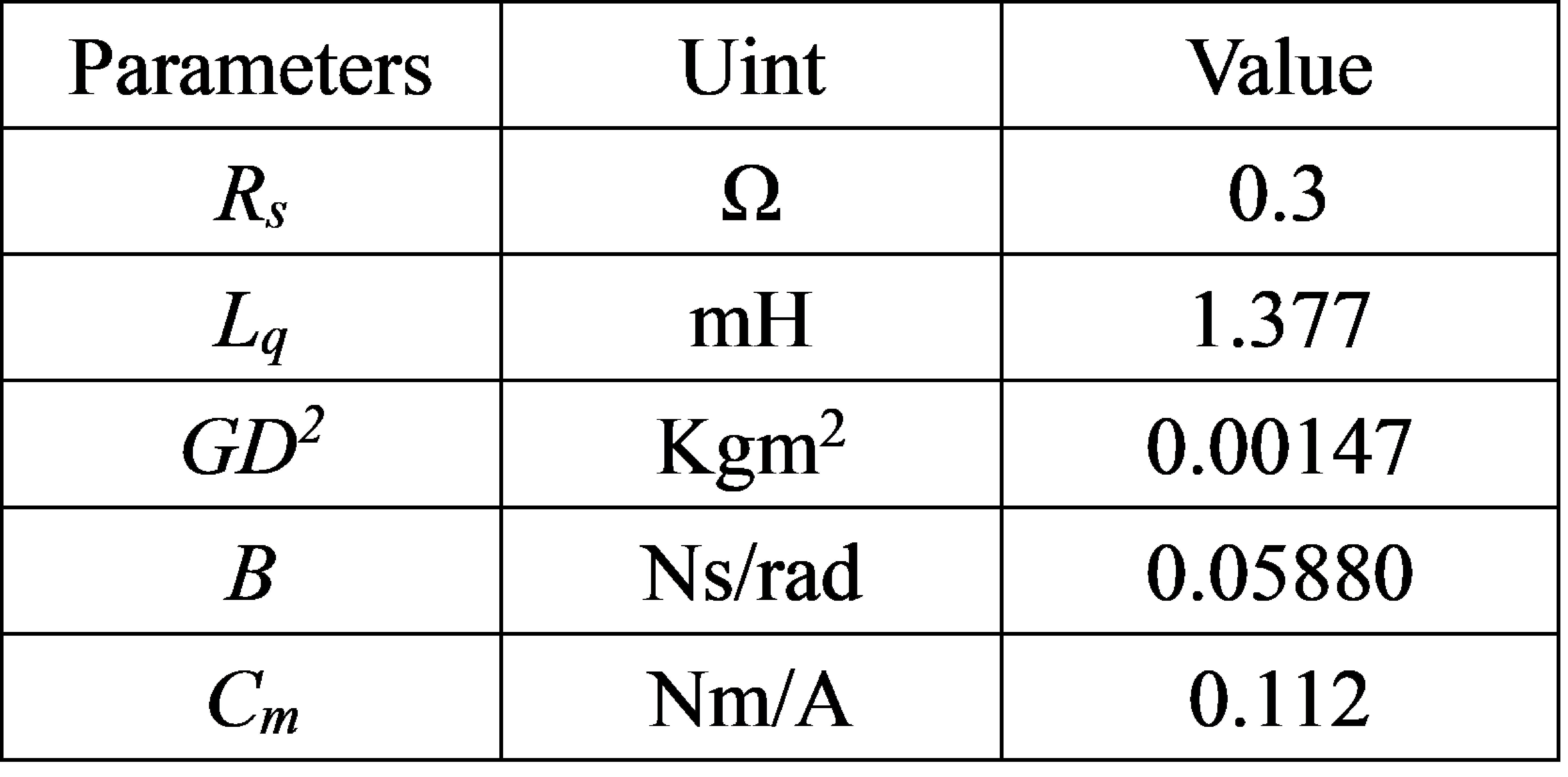}
		\centering
		\label{table_3}
	\end{table}
}

When the specification of the PMSM in Table \ref{table_3}  is used,  the   plant  (\ref{eq_45}) of the PMSM becomes
\begin{equation}
G(s) = \frac{{2380.9}}{{{s^2} + 138.1s + 3819.7}}
\end{equation}
We use the same controllers in Section \ref{sec:TDS} for IFO-ADRC, FO-ADRC and IO-ADRC. Let the observer gain be $L=[\beta_1,\beta_2,\beta_3]\t=[3\omega_0,3\omega_0^2,\omega_0^3]\t$, ${\omega _o} = 700 $ rad/s, and $\gamma=0.75$. 
The operator ${s^\gamma }$ is discretized by the impulse response invariant method where
the discrete frequency  is 8000 Hz and the discrete order of the fractional-order operators is 5.  The parameter of the P controller (see (\ref{eq:CP})) used in the IFO-ADRC and the FO-ADRC is $K_{fp} = 750$, while
the PD controller used in the IO-ADRC design method in Section \ref{sec:TDS} is designed as
\begin{equation}
{C_{pd}}(s) = 2314.69(1 + 0.0185s)
\end{equation}

For the IFO-ADRC closed-loop system, all parameters meet the conditions of Theorem \ref{th_closed-loop}, thus the closed-loop  of IFO-ADRC is BIBO stable. Also, the IO-ADRC and FO-ADRC closed-loop systems are ensured to be BIBO stable.

\begin{figure}[!ht]
	\centering
	\includegraphics[scale=0.50]{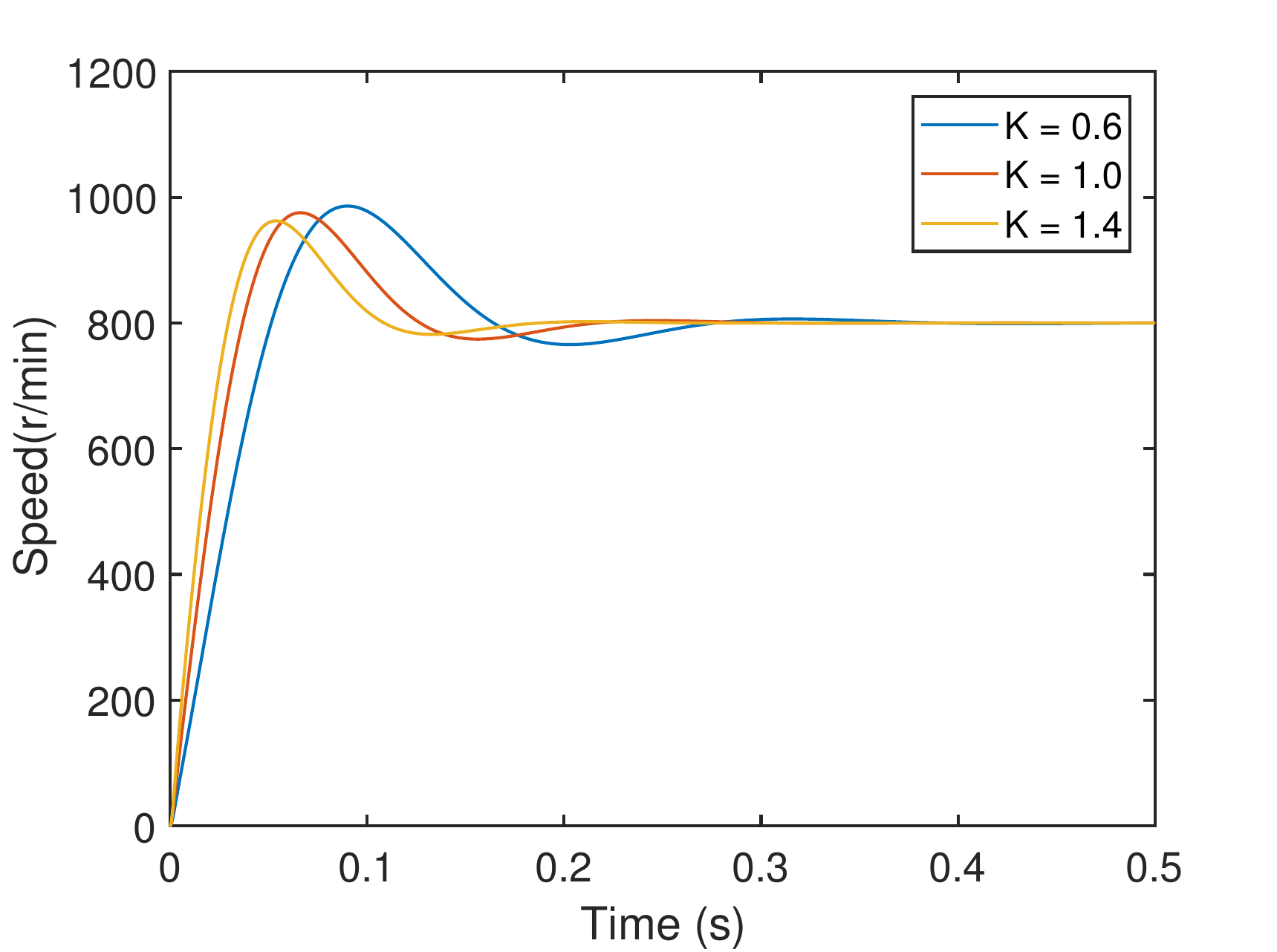}
	\caption{Step responses of the IO-ADRC  control system with controller gain variations (simulation)}
	\label{figure_11}
\end{figure}

\begin{figure}[t]
	\centering
	\includegraphics[scale=0.50]{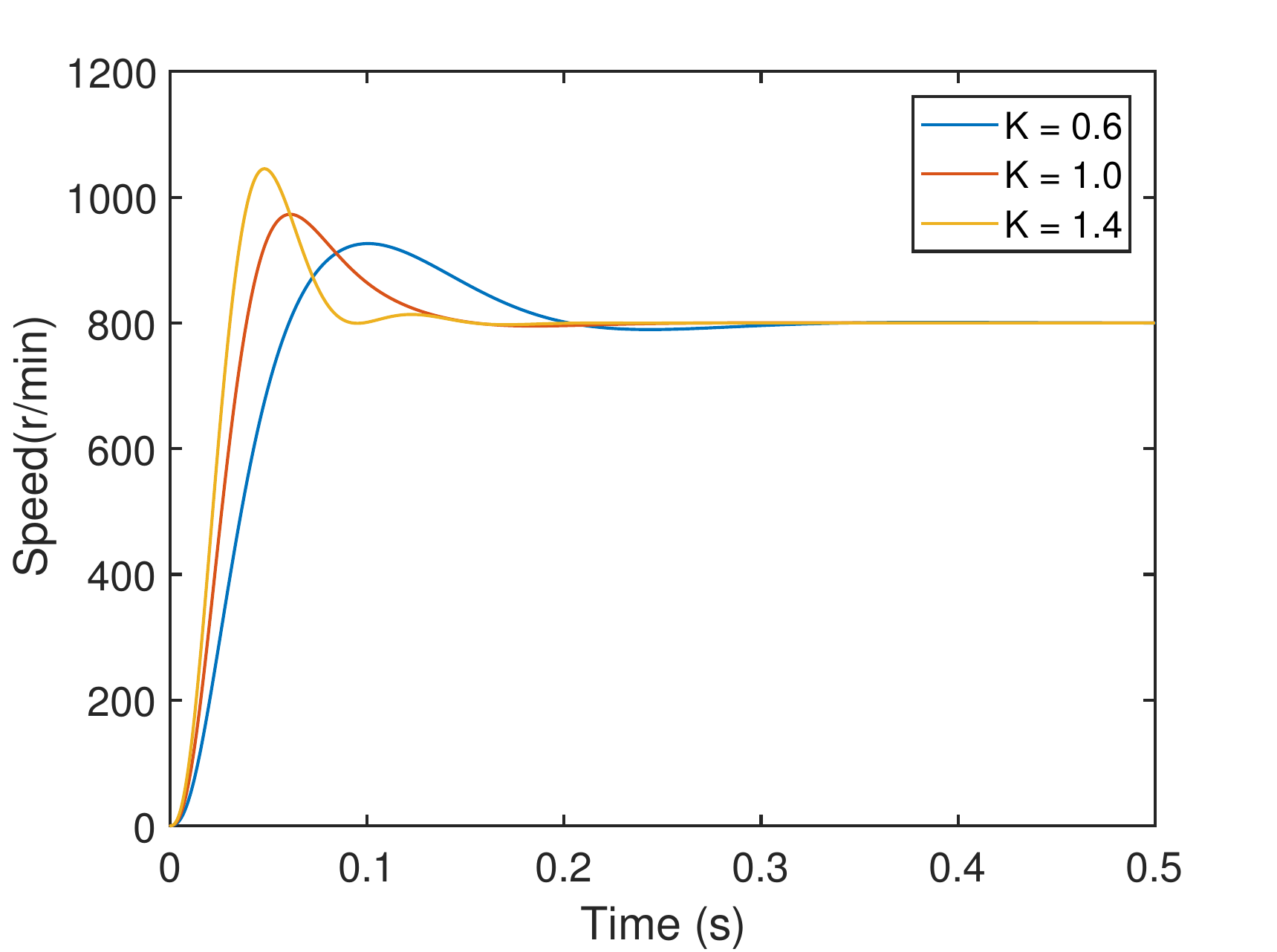}
	\caption{Step responses of the FO-ADRC  control system with controller gain variations (simulation)}
	\label{figure_12}
\end{figure}

\begin{figure}[t]
	\centering
	\includegraphics[scale=0.50]{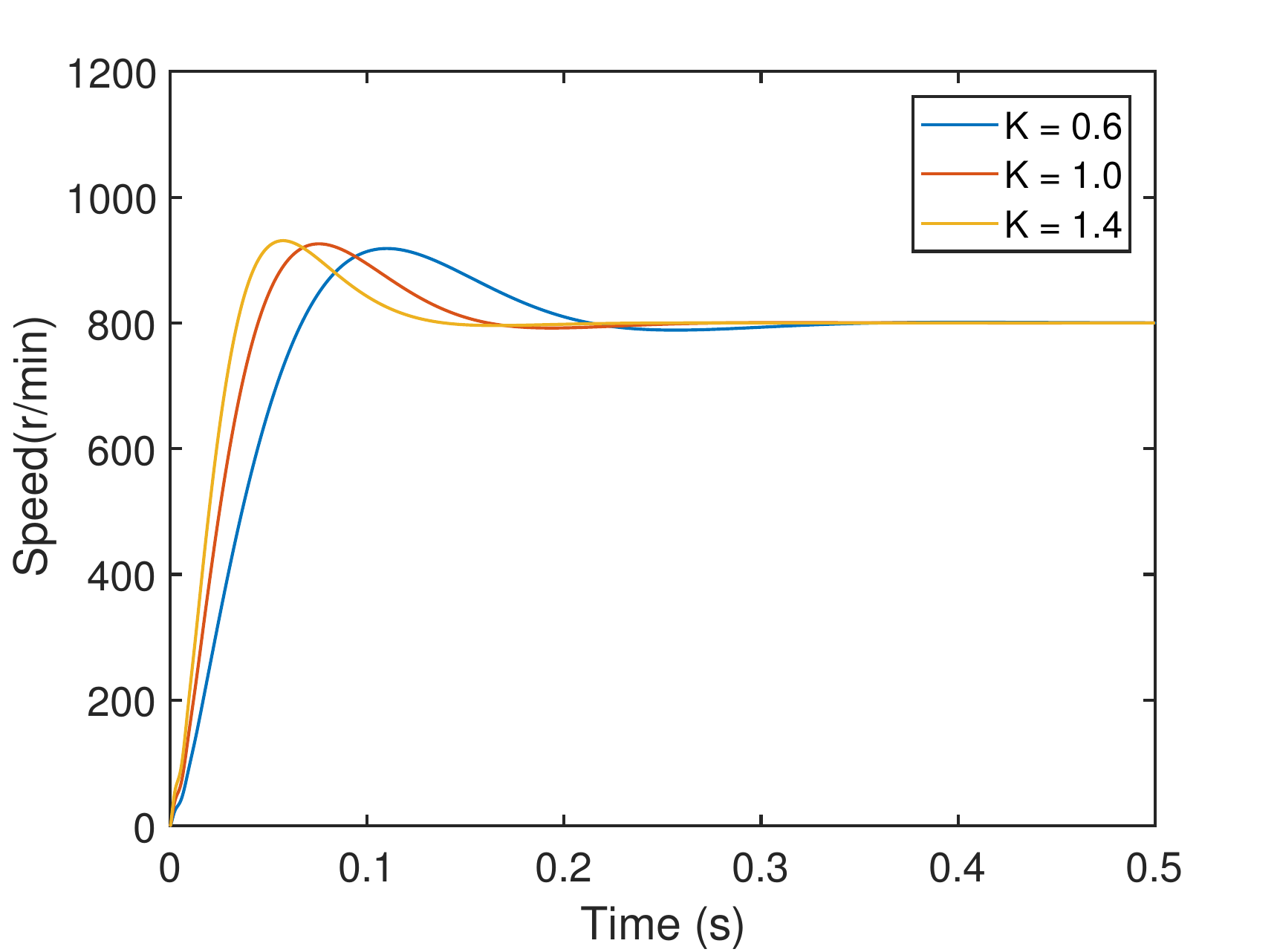}
	\caption{Step responses of the IFO-ADRC  control systemwith controller gain variations (simulation)}
	\label{figure_14}
\end{figure}

\begin{figure}[!ht]
	\centering
	\includegraphics[scale=0.50]{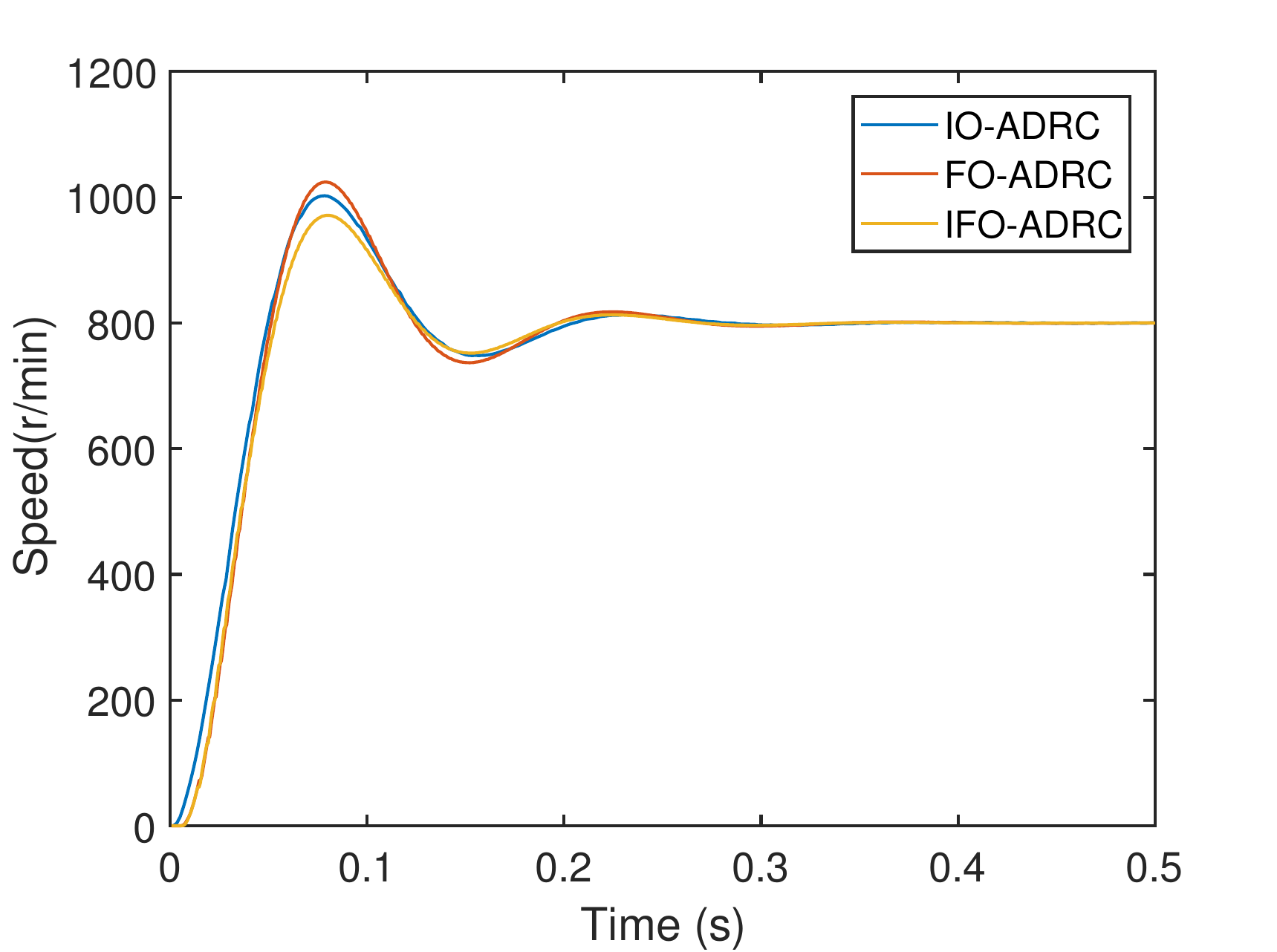}
	\caption{Step responses of three different control methods (experiment)}
	\label{figure_23}
\end{figure}

\begin{figure}[!ht]
	\centering
	\includegraphics[scale=0.50]{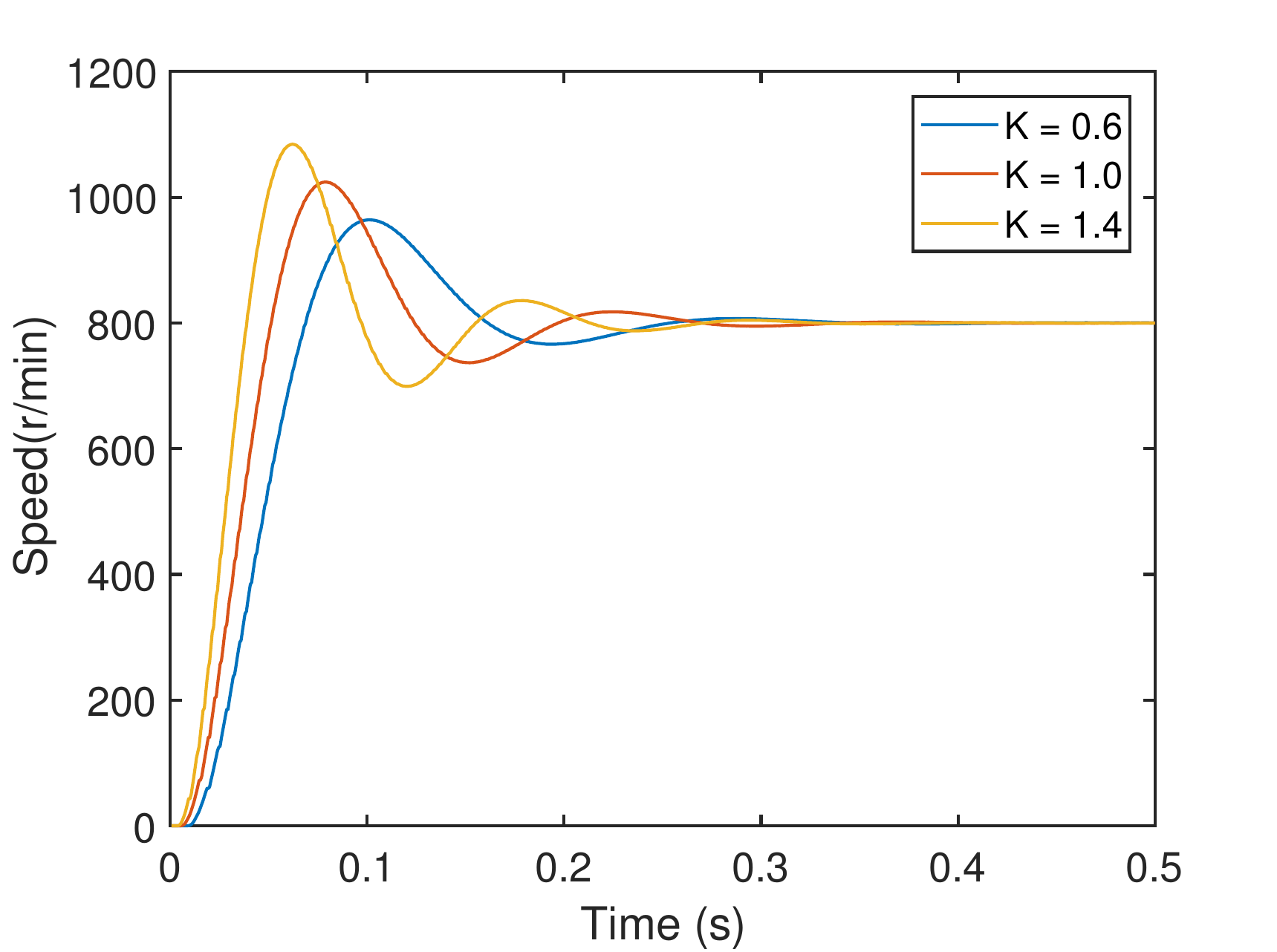}
	\caption{Step responses of the FO-ADRC with controller gain variations (experiment)}
	\label{figure_16}
\end{figure}

\begin{figure}[!ht]
	\centering
	\includegraphics[scale=0.50]{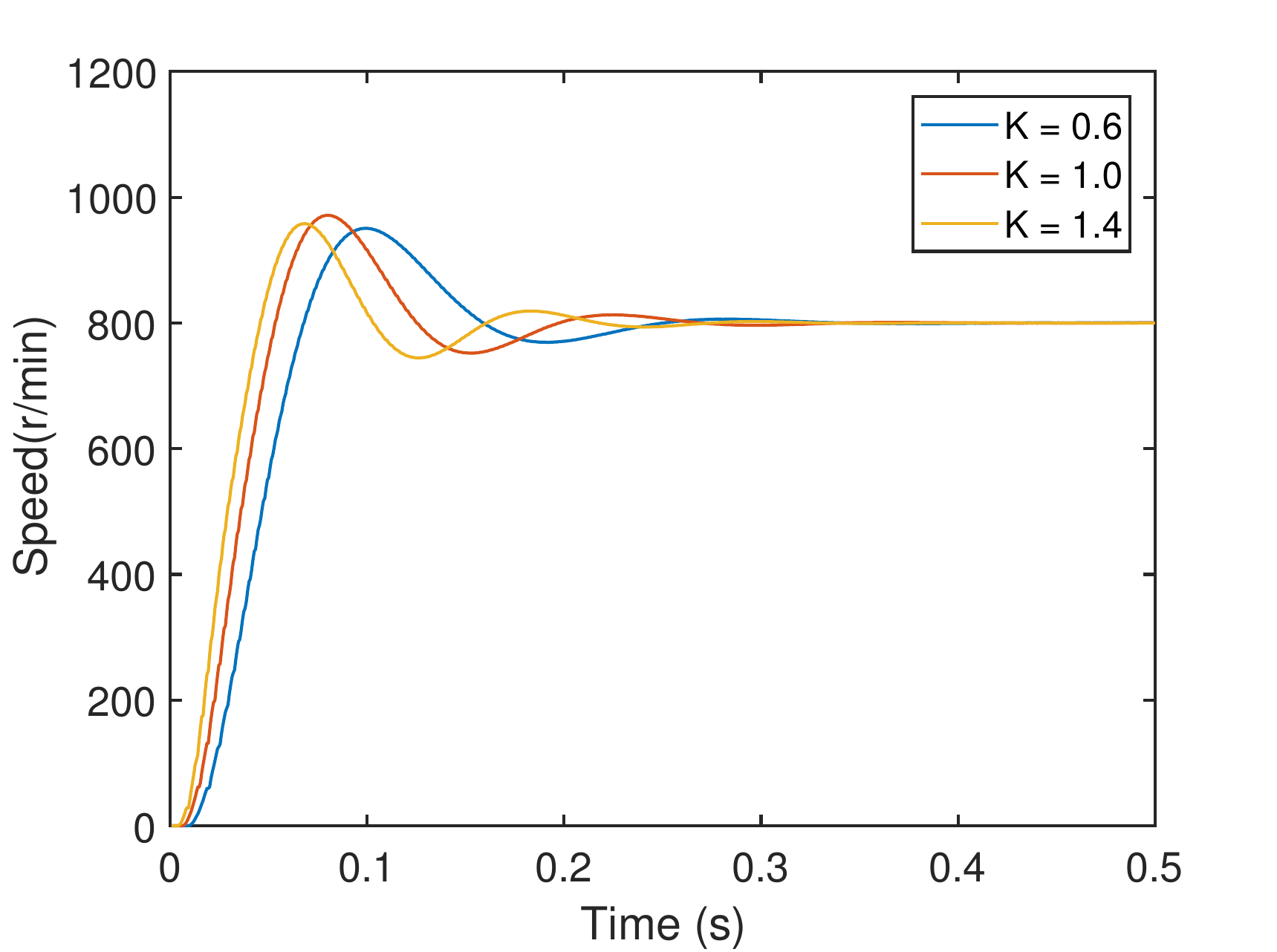}
	\caption{Step responses of the IFO-ADRC with controller gain variations (experiment)}
	\label{figure_17}
\end{figure}

Let us multiple the P controller parameter $K_{fp}$ and PD controller parameter 
	$K_{ip}$ by $K=0.6$ and $K=1.4$, while the nominal value is used when $K=1$.
	Fig.~\ref{figure_12} and Fig.~\ref{figure_14} are simulation results of step responses with different controller parameters for the FO-ADRC and IFO-ADRC systems, respectively.
	As shown in Fig.~\ref{figure_12} and Fig.~\ref{figure_14}, the IFO-ADRC system is robust to the controller parameter variations. Fig.~\ref{figure_23} is step responses of the IO-ADRC, FO-ADRC and IFO-ADRC systems for the experiment setup. It is illustrated that the step response of the IFO-ADRC system has smaller overshoot, oscillation magnitude and shorter settling time than the FO-ADRC and IO-ADRC system. Fig.~\ref{figure_16} and Fig.~\ref{figure_17} are experiment results of step responses  of three different systems  when the controller parameter varies. Similar to simulation result, the IFO-ADRC system is the most robust to controller parameter variations.  Table \ref{table_2} summarizes  results of the step responses for three different systems, showing that the IFO-ADRC system has smaller overshoot, less settling time, and overshoot flutuation than the IO-ADRC and IFO-ADRC system.

\begin{figure}[!ht]
	\centering
	\includegraphics[scale=0.50]{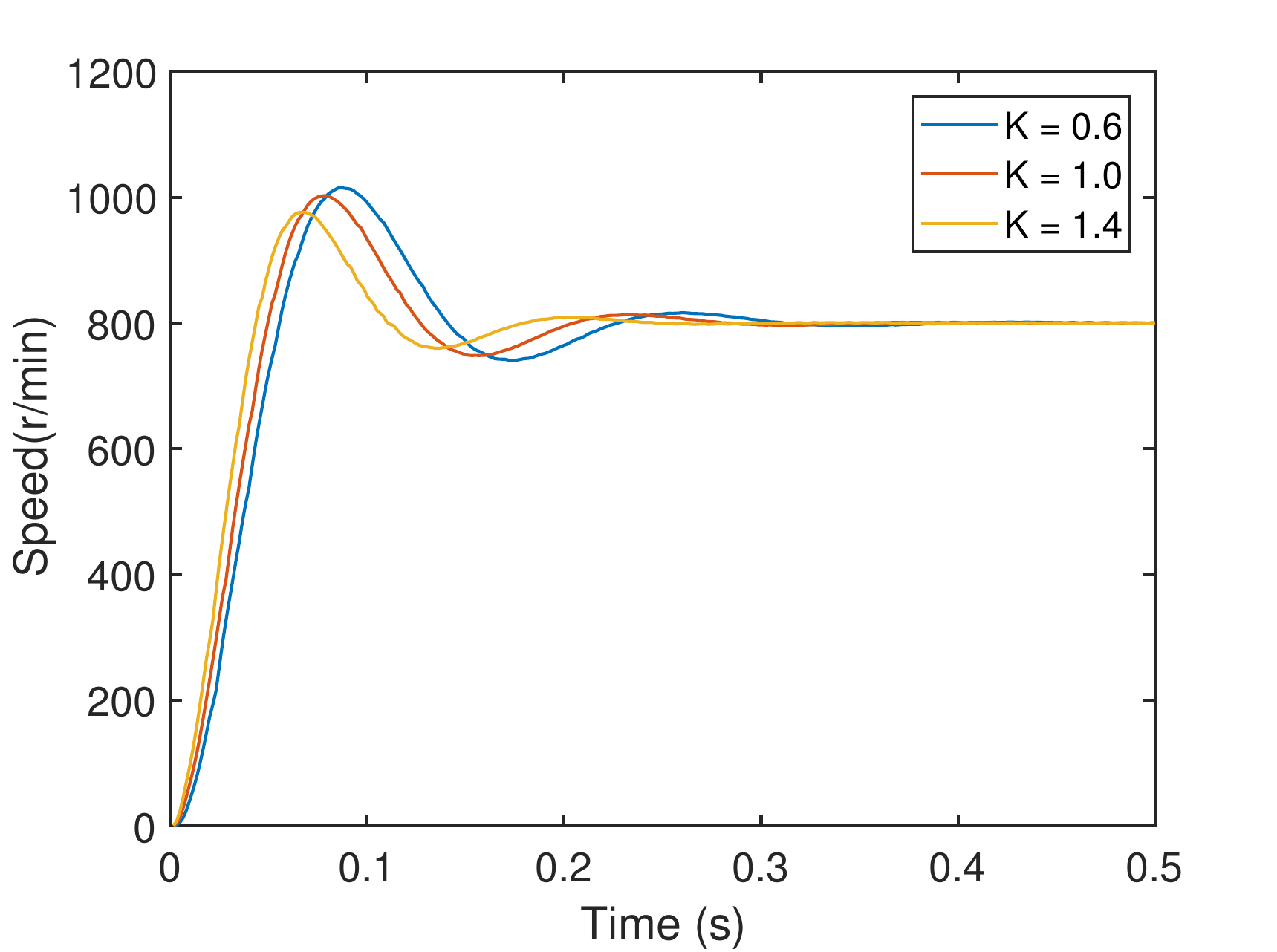}
	\caption{Step responses of the IO-ADRC with controller gain variations (experiment)}
	\label{figure_19}
\end{figure}

\begin{table}[!ht]
	\renewcommand{\arraystretch}{1.3}
	\caption{Comparison of the responses with three control systems (experiment) }
	\centering
	\includegraphics[scale=0.080]{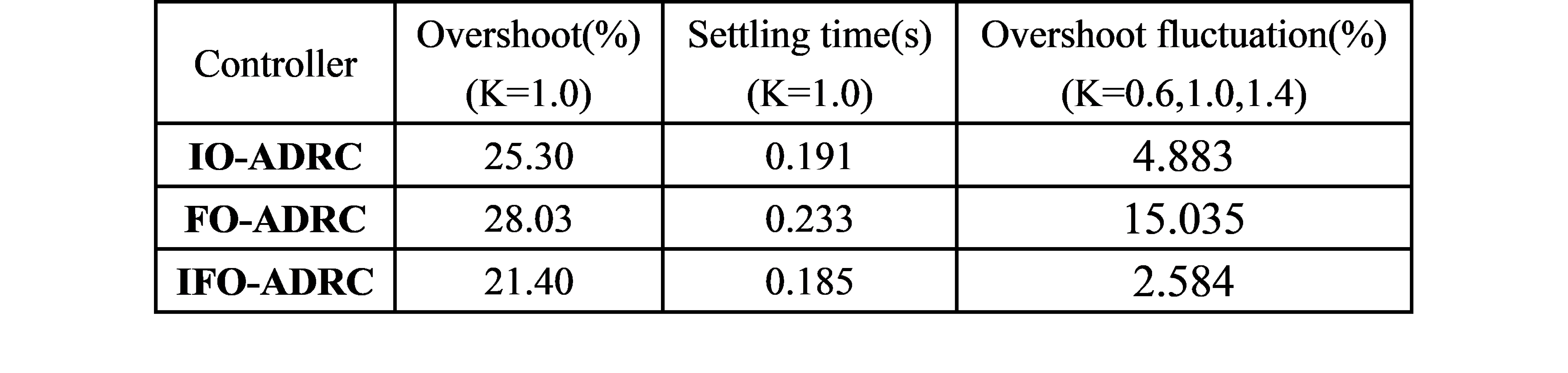}
	\label{table_2}
\end{table}

\section{Conclusion}
An improved active disturbance rejection control scheme with IFO-ESO is proposed in this paper. IFO-ESO can approximately convert an integer-order system into a cascaded  fractional-order integrator for which a simpler feedback law can be designed.
The approximation to a cascaded fractional-order integrator by IFO-ESO behaves well
across  a larger frequency band than FO-ESO, ensuring the closed-loop system is robust to controller gain, ESO parameters, and plant parameters variations.  
The frequency-domain analysis and PMSM speed servo control experiments verify that the proposed IFO-ADRC achieves better  performance than FO-ADRC and IO-ADRC. 

\appendix
\prooflater{Theorem \ref{th_IFO-ESO}}
Let $w = s^{\nu + \gamma}$, and (\ref{eq_32_}) can be written:
\begin{equation}
\lambda (w) = w({w^{n\sigma }} + \sum\limits_{i = 1}^{n - 1} {{\beta _i}} {w^{(n - i)\sigma }}) + {\beta _n}{w^\sigma } + {\beta _{n + 1}}
\label{eq_33_}
\end{equation}
where $\sigma  = \frac{\gamma }{{\nu  + \gamma }}$, $0 < \sigma  < 1$. 
According to Kharitonov-Based Method \cite{petravs2011fractional}, the two boundary polynomials are:
\begin{equation}
\begin{array}{l}
{}^1\lambda (w) = w(1 + \sum\limits_{i = 1}^{n - 1} {{\beta _i}} ) + {\beta _n} + {\beta _{n + 1}}\\
{}^2\lambda (w) = w({w^n} + \sum\limits_{i = 1}^{n - 1} {{\beta _i}} {w^{(n - i)}}) + {\beta _n}w + {\beta _{n + 1}}
\end{array}
\label{eq_34_}
\end{equation}
Substitute ${\beta _i} = C_{n + 1}^i{\omega _o}^i$, $i = 1,2, \cdots ,n+1$ into ${}^2\lambda (w)$ gives ${}^2\lambda (w) = {(w + {\omega _o})^{n + 1}}$. Then,  the roots of the two boundary polynomials are:
\begin{equation}
\begin{array}{l}
{}^1\lambda (w) : {w_1} =  - \frac{{n{\omega _o}^n + {\omega _o}^{n + 1}}}{{(1 + \sum\limits_{i = 1}^{n - 1} {C_{n + 1}^i} {\omega _o}^i)}}\\
{}^2\lambda (w) : {w_i} =  - {\omega _o}, i= {1,2,3 \cdots ,n + 1}
\end{array}
\label{eq_35_}
\end{equation}
Since all the roots of the boundary polynomials are loacted in $|\mbox{arg}({w_i})| > \frac{\pi }{{2(\nu  + \gamma )}}$, all the roots of (\ref{eq_33_}) are loacted in $|\mbox{arg}({w_i})| > \frac{\pi }{{2(\nu  + \gamma )}}$ \cite{petravs2002robust}, i.e., $\lambda (s)$ is Hurwitz, system (\ref{eq_30_}) is BIBO stable, regarding $h_{ifo}$ as input and $e_1$ as output.\eproof

\prooflater{Theorem \ref{th_closed-loop}}

Let $({r_1},{r_2},{r_3} \cdots {r_{m}},{r_{m+1}}) = (r,r^{(1)}, r^{(2)}, \cdots ,{r^{(m-1)}},{r^{(n\gamma-m+1)}})$. From (\ref{eq_14}) and (\ref{eq_30_}), one has
\begin{gather}
{y^{(n\gamma )}} = {k_p}({r_1} - {z_1}) + {k_{{d_1}}}({\dot{r}_1} - {\dot {z}_1}) +  \ldots+ {k_{{d_{m - 2}}}}({r_1}^{(m-2)}  \nonumber \\
- {z_1}^{(m-2)})+ {r_{1}^{(n\gamma)}} + {f_{ifo}} - {{\mathord{\buildrel{\lower3pt\hbox{$\scriptscriptstyle\frown$}} 
			\over f} }_{ifo}} + q - \mathord{\buildrel{\lower3pt\hbox{$\scriptscriptstyle\frown$}} 
	\over q} 
\end{gather}
It follows from (\ref{eq:ei}) that
\begin{gather}
{y^{(n\gamma )}} - {r_1}^{(n\gamma )} = {k_p}({r_1} - {x_1} + {e_1}) + {k_{{d_1}}}({{\dot r}_1} - {{\dot x}_1} + {{\dot e}_1}) +  \ldots \nonumber\\
+ {k_{{d_{m - 2}}}}({r_1}^{(m - 2)} - {x_1}^{(m - 2)} + {e_1}^{(m - 2)})\nonumber\\
+ {e_{n + 1}} + {e_n}^{(\gamma )} - {e_n}^{(\chi )}
\label{eq_41_}
\end{gather}

\begin{figure}[!ht]
		\centering
		\includegraphics[scale=0.045]{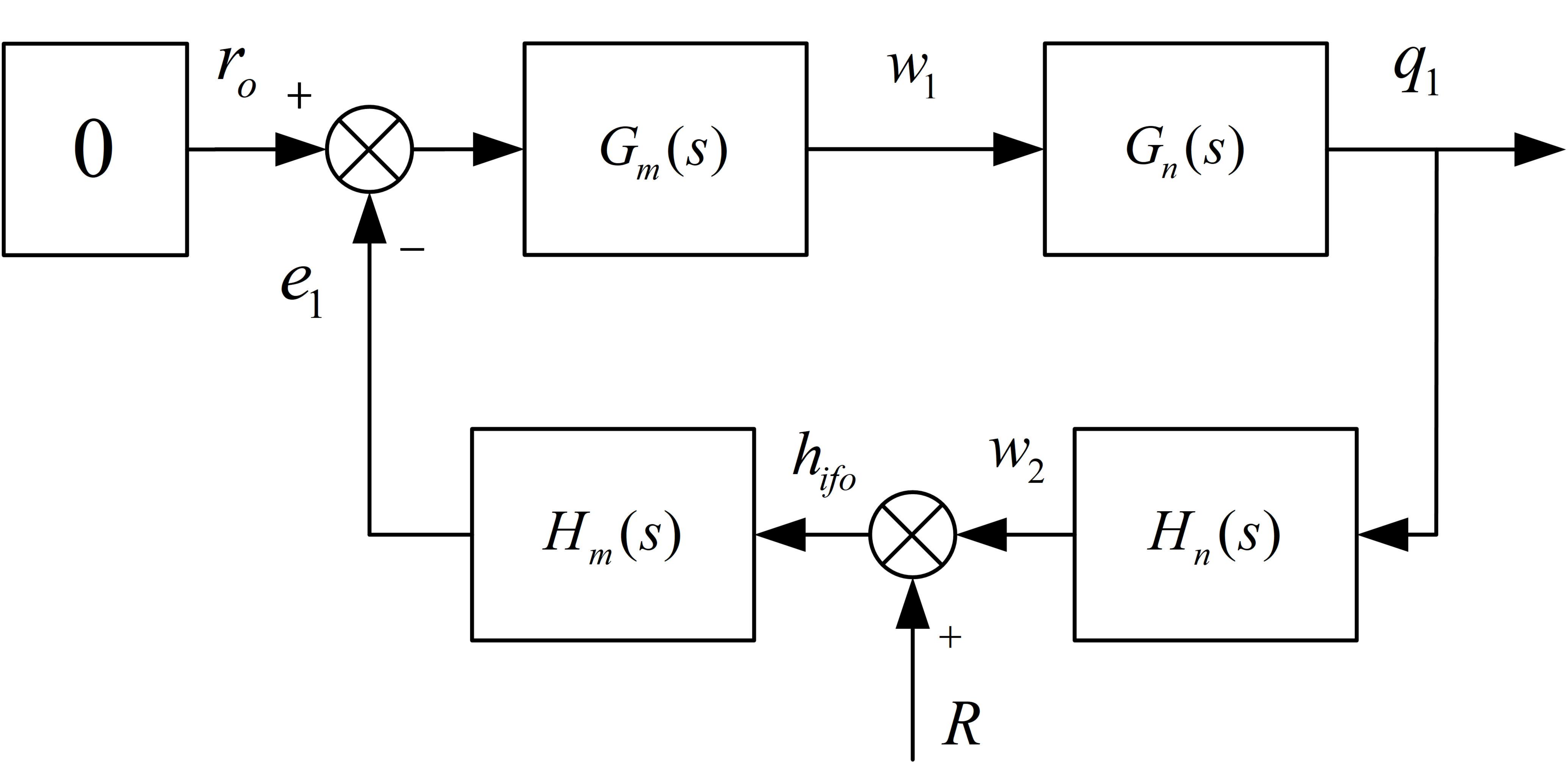}
		\caption{The block digram of the closed-loop system}
		\label{figure_24}
	\end{figure}

Let ${q_1} = {r_1} - {x_1}$, ${q_i} = {\dot{q}_{i - 1}}, i = 2,3, \cdots ,v-1$, (\ref{eq_41_}) can be written as
\begin{gather}
{{\dot q}_1} = {q_2},\;
{{\dot q}_2} = {q_3},\;
\cdots,\;
{{\dot q}_{m - 1}} = {q_m}\nonumber\\
{q_m}^{(n\gamma  - m + 1)} = {q_{m + 1}} =  - ({k_p}{q_1} + {k_{{d_1}}}{q_2} +  \cdots  + {k_{{d_{m - 2}}}}{q_{m-1}})\nonumber\\
- ({k_p}{e_1} + {k_{{d_1}}}{{\dot e}_1} +  \cdots  + {k_{{d_{m - 2}}}}{e_1}^{(m - 2)})\nonumber\\
- {e_{n + 1}} - {e_n}^{(\gamma )} + {e_n}^{(\chi )}
\label{eq-62}
\end{gather}
The system (\ref{eq_30_}) can be written:
\begin{align}
{e_1}^{(\gamma )} =&  - {\beta _1}{e_1} + {e_2} \nonumber \\
{e_2}^{(\gamma )} =&  - {\beta _2}{e_1} + {e_3}\nonumber\\
&\vdots \nonumber\\
\frac{{{d^{(\chi )}}e_n}}{{{d^{(\chi )}}t}}=&   - {\beta _{n}}{e_1} + {e_{n + 1}}\nonumber\\
{e_{n + 1}}^{(\gamma )} =&  - {\beta _{n+1}}{e_1} + {h_{ifo}}\nonumber\\
{h_{ifo}} =& {a_0}{q_1}^{(\gamma )} + {a_1}{q_2}^{(\gamma )} + {a_2}{q_3}^{(\gamma )} \nonumber \\
&+  \cdots  + {a_{m - 1}}{q_m}^{(\gamma )} + R\nonumber
\end{align} 
where
\begin{equation}
R =  - \sum\limits_{i = 1}^{n - 1} {{a_i}} {r_{i + 1}}^{(\gamma )} - {a_0}{r_1}^{(\gamma )} + d 
\label{eq-63}
\end{equation}
	According to (\ref{eq-62}), it follows that,
\begin{align} \label{eq:new_62}
{q_m}^{(n\gamma  - m + 1)} =& {q_{m + 1}} =  - ({k_p}{q_1} + {k_{{d_1}}}{q_2} +  \cdots  + {k_{{d_{m - 2}}}}{q_{m-1}})\nonumber\\
&- ({k_p}{e_1} + {k_{{d_1}}}{{\dot e}_1} +  \cdots  + {k_{{d_{m - 2}}}}{e_1}^{(m - 2)})\nonumber \\
& - {e_n}^{(\gamma )} -{\beta_n}{e_1}
\end{align} 
Note ${w_1}$ = ${k_p}{e_1} + {k_{{d_1}}}{\dot e_1} +  \cdots  + {k_{{d_{m - 2}}}}{e_1}^{(m - 2)} + {e_n}^{(\gamma )} + {\beta_n}{e_1}$,${w_2} = {a_0}{q_1}^{(\gamma )} + {a_1}{q_2}^{(\gamma )} + {a_2}{q_3}^{(\gamma )} +  \cdots  + {a_{m - 1}}{q_m}^{(\gamma )}$. Combining (\ref{eq:new_62}) and (\ref{eq-63}) gives the block diagram of the closed-loop system in Fig. \ref{figure_24}.
From (\ref{eq:new_62}), it gives
\begin{align}
{G_m}(s) =& -\frac{{{W_1}(s)}}{{E(s)}} = {k_p} + \sum\limits_{i = 1}^{m - 2} {{k_{{d_i}}}{s^i}}  + {s^{n\gamma }} + \sum\limits_{i = 1}^n {{\beta _i}{s^{(n - i)\gamma }}}\nonumber \\
{G_n}(s) =& \frac{{{Q_1}(s)}}{{{W_1}(s)}} = \frac{1}{{{s^{n\gamma }} + {k_p} + \sum\limits_{i = 1}^{m - 2} {{k_{{d_i}}}{s^i}} }}
\end{align}
where $W_1(s)$, $Q_1(s)$ and $E_1(s)$ are the Laplace transforms of signals $w_1$, $q_1$ and $e_1$, respectively.
Accoding to (\ref{eq-63}), one has
\begin{align}
{H_m}(s) =& \frac{{{E_1}(s)}}{{{H_{f}}(s)}} = \frac{1}{{{s^{\nu  + \gamma }}({s^{n\gamma }} + \sum\limits_{i = 1}^{n - 1} {{\beta _i}} {s^{(n - i)\gamma }}) + {\beta _n}{s^\gamma } + {\beta _{n + 1}}}}\nonumber \\
{H_n}(s) =& \frac{{{W_2}(s)}}{{{Q_1}(s)}} = {s^r}\sum\limits_{i = 0}^{m - 1} {{a_i}{s^i}}     
\end{align}
where $H_f(s)$ and $W_2(s)$ are the Laplace transforms of signals $h_{ifo}$ and $w_2$, respectively.    
    
Now, let us show $R$ is bounded. Let ${d_0} = \sum\limits_{i = 0}^{m - 2} {C_m^i{v^{m - i}}{r_{i+1}}}  + {r_{m + 1}}$ where $v>0$ and $d_1 = \sum\limits_{i = 1}^{n - 1} {{a_i}} {r_{i + 1}}^{(\gamma )} + {a_0}{r_1}^{(\gamma )} $. Define the transfer function of the system as
\begin{gather}
{P_1}(s) = \frac{{{D_1}(s)}}{{{D_0}(s)}} = \frac{{\sum\limits_{i = 1}^{n - 1} {{a_i}} {s^{(i + \gamma )}} + {a_0}{s^{(\gamma )}}}}{{{s^{n\gamma }} + \sum\limits_{i = 0}^{m - 2} {C_m^i{v^{m - i}}{s^i}} }}
\end{gather}
where $D_0(s)$ and $D_1(s)$ are the Laplace transforms of $d_0$ and $d_1$, respectively.
Similar to the proof of (\ref{eq_32_}), the characteristic polynominal of the system $P_1(s)$ is Hurwitz, ie., the system $P_1(s)$ is BIBO stable. As a result, ${r_1},{r_2},{r_3} \cdots {r_{m}},{r_{m+1}}$ are  bounded, i.e., $d_0$ is bounded, then $d_1$ is bounded. Note that $R$ = $-d_1+d$. Since $d$ is bounded,  $R$ is bounded.

Because $R$ is bounded, we can treat $R$ as the disturbance and calculate the transfer function of the closed-loop system. Further calculation gives    
\begin{align}
{G_o}{\rm{(s)  =  - }}\frac{{{Q_1}(s)}}{{E(s)}}{\rm{ = }}{{\rm{G}}_m}(s){G_n}(s)
\label{eq-64}
\end{align}
\begin{align}
{H_o}(s) = \frac{{{E_1}(s)}}{{{Q_1}(s)}}{\rm{ = }}{H_m}(s){H_n}(s)
\label{eq-65}
\end{align}

Combining (\ref{eq-64}) and (\ref{eq-65}), the transfer function of the closed-loop system $P_o(s)$ can be given 
\begin{gather}
{P_o}(s) = \frac{{{Q_1}(s)}}{{{R_o}(s)}} = \frac{{{G_o}(s)}}{{1 + {G_o}(s){H_o}(s)}}
\end{gather}
where $R_o(s)$ is the Laplace transfoms of $r_o$ (see the Fig. \ref{figure_24}).
The characteristic polynomial of the closed-loop system is
\begin{align}
P(s) =& ({s^r}\sum\limits_{i = 0}^{m - 1} {{a_i}{s^i}}) ({k_p} + \sum\limits_{i = 1}^{m - 2} {{k_{{d_i}}}{s^i}}  + {s^{n\gamma }} + \sum\limits_{i = 1}^n {{\beta _i}{s^{(n - i)\gamma }}} )\nonumber \\
+& (s^{n\gamma}+{k_p} + \sum\limits_{i = 1}^{m - 2} {{k_{{d_i}}}{s^i}} )({s^{\nu  + \gamma }}({s^{n\gamma }} + \sum\limits_{i = 1}^{n - 1} {{\beta _i}} {s^{(n - i)\gamma }})\nonumber \\ +& {\beta _n}{s^\gamma } + {\beta _{n + 1}})
\end{align}
Finding  prime number $p_1$, $p_2$, $q_1$, and $q_3$  such that    $\nu = \frac{p_1}{q_1}$, $\gamma = \frac{p_2}{q_2}$, (\ref{eq-19}) is satisfied. Since $R$ is bounded, the closed-loop system is BIBO stable \cite{petravs2011fractional}. 
\eproof

\prooflater{Proposition \ref{th_closed-loop}}
When $m=n=2$, the characteristic polynomial $P(s)$ of the closed-loop system can be written
\begin{align}
P(s) =& {{s^r}({a_0} + {a_1}s)({k_p} + {s^{2\gamma }} + {\beta _1}{s^\gamma } + {\beta _2})}\\
+& {(s^{2\gamma}+{k_p})({s^{2 + \gamma }}+{\beta _1}{s^2} + {\beta _2}{s^\gamma } + {\beta _3})}
\end{align}
According to Kharitonov-Based Method, when ${\beta _i} = C_{n + 1}^i{\omega _o}^i$ for $i = 1,2,3$, the two boundary polynomial can be written
\begin{align}
{}^1P(s) =& {{A_0}{s^2} + {A_1}s + {A_2}}\nonumber \\
{}^2P(s) =& {{B_0}{s^5} + {B_1}{s^4} + {B_2}{s^3} + {B_3}{s^2} + {B_4}s + {B_5}}
\end{align}
where
\begin{align}
{A_0} =& {1 + k_p + 3(1 + k_p){\omega _o}} \nonumber\\
{A_1} =& {{a_1}( 1 + k_p + 3{\omega _o} + 3{\omega _o}^2)} \nonumber \\
{A_2} =& {{a_0}( 1 + k_p + 3{\omega _o} + 3{\omega _o}^2) + 3(1 + k_p){\omega _o}^2}\nonumber \\ +& {(1 + k_p){\omega _o}^3}\nonumber \\
{B_0} =& 1,\; 
{B_1} =  {{a_1} + 3{\omega _o}},\;
{B_2} =  {{a_0} + k_p + 3{a_1}{\omega _o} + 3{\omega _o}^2}\nonumber \\
{B_3} =& {{a_1}k_p + 3{a_0}{\omega _o} + 3k_p{\omega _o} + 3{a_1}{\omega _o}^2 + {\omega _o}^3}\nonumber \\
{B_4} =& {{a_0}k_p + 3{a_0}{\omega _o}^2 + 3k_p{\omega _o}^2},\;
{B_5} = {k_p{\omega _o}^3}
\label{eq-69}
\end{align}

Firstly, we consider ${}^1P(s) = 0$. When $a_1 \ge 0$, $a_0 \ge 0$, $k_p>0$, and $\omega_o > 0$, from (\ref{eq-69}), we have $A_0>0$, $A_1 \ge 0$, and $A_2 > 0$. 
Accoring to Routh-Hurwitz criterion, if $a_1 \ge 0$, $a_0 \ge 0$, $k_p>0$, and $\omega_o>0$, all the roots of ${}^1P(s)=0$ are located in left plane, i.e., when $a_1 \ge 0$ and ${a_0} \ge 0$, there exists a sufficiently large  $\omega_o > 0$, such that all the root of ${}^1P(s)=0$ are located in left plane.

Secondly, we consider ${}^2P(s) = 0$. According to Routh-Hurwitz criterion, ${}^2P(s) = 0$ can be written as the Routh table form 
\begin{table}[!ht]
	\renewcommand{\arraystretch}{1.3}
	\caption{Routh table of ${}^2P(s) = 0$}
	\centering
	\includegraphics[scale=0.045]{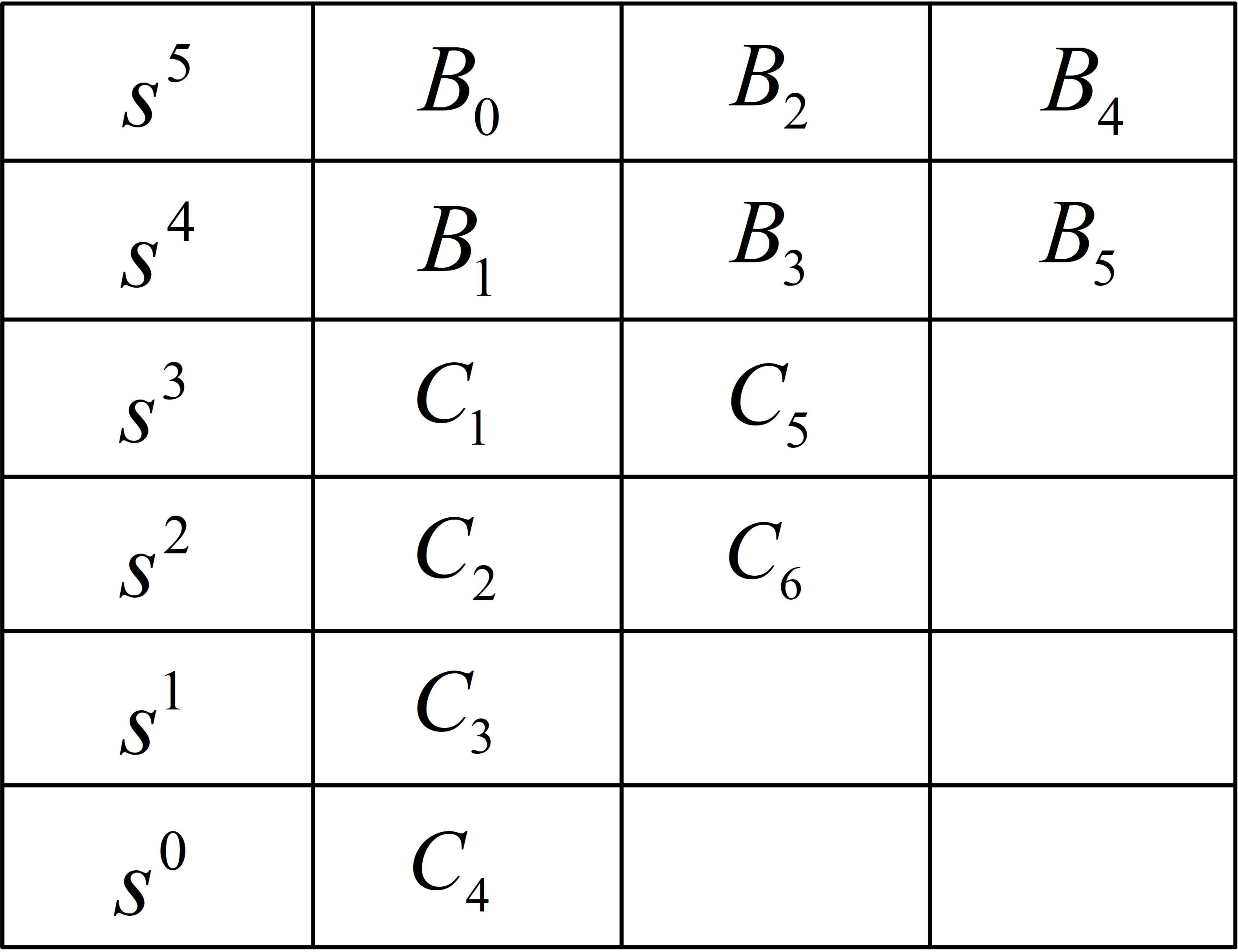}
	\label{table_4}
\end{table}
\\
In TABLE \ref{table_4}, $C_1$, $C_2$, $C_3$, and $C_4$ are
\begin{equation}
{C_1} = \frac{{{N_5}}}{{{D_5}}},\;
{C_2} = \frac{{{N_6}}}{{{D_6}}},\;
{C_3} = \frac{{{N_7}}}{{{D_7}}},\;
{C_4} = k_p{\omega _o}^3
\end{equation}
where
\begin{align}
{N_5} =& {{a_0}{a_1} + 3{a_1}^2{\omega _o} + 9{a_1}{\omega _o}^2 + 8{\omega _o}^3} \nonumber \\
{D_5} =& {{a_1} + 3{\omega _o}}\nonumber \\
{N_6} =& 3{a_1}({a_0}^2 + {a_1}^2{k_p}){\omega _o} + (9{a_0}{a_1}^2 + 15{a_1}^2{k_p}){\omega _o}^2\nonumber \\ +& {a_1}(10{a_0} + 9{a_1}^2 + 18{k_p}){\omega _o}^3 + 30{a_1}^2{\omega _o}^4 + 33{a_1}{\omega _o}^5\nonumber \\ +& 8{\omega _o}^6 - 3{a_1}{a_0}{k_p} - 9{a_0}{k_p} - 3{a_0}{\omega _o}^4\nonumber \\
{D_6} =& {{a_0}{a_1} + 3{a_1}^2{\omega _o} + 9{a_1}{\omega _o}^2 + 8{\omega _o}^3\nonumber} \\
{N_7} =& 3({a_0}^3{a_1}{k_p} + {a_0}{a_1}^3{k_p}^2)+ (9{a_0}^2{a_1}^2{k_p} \nonumber \\ +& 15{a_0}{a_1}^2{k_p}^2){\omega _o} + (9{a_0}^3{a_1} + 9{a_0}^2{a_1}{k_p}\nonumber \\  +& 18{a_0}{a_1}^3{k_p} + 9{a_0}{a_1}{k_p}^2 + 9{a_1}^3{k_p}^2){\omega _o}^2 + (27{a_0}^2{a_1}^2\nonumber \\+& 96{a_0}{a_1}^2{k_p} - 24{a_0}{k_p}^2 + 42{a_1}^2{k_p}^2){\omega _o}^3 + (30{a_0}^2{a_1} \nonumber \\ +& 27{a_0}{a_1}^3 + 108{a_0}{a_1}{k_p} + 18{a_1}^3{k_p} + 48{a_1}{k_p}^2){\omega _o}^4 \nonumber \\ +& (90{a_0}{a_1}^2 + 54{a_1}^2{k_p}){\omega _o}^5 + (99{a_0}{a_1} + 48{a_1}{k_p}){\omega _o}^6\nonumber \\ +& 24{a_0}{\omega _o}^7 - 3{a_0}^2{a_1}{k_p}^2 - 9{a_0}^2{k_p}^2{\omega _o} - 30{a_0}^2{k_p}{\omega _o}^3 \nonumber \\-& 9{a_0}^2{\omega _o}^5 \nonumber \\
{D_7} =&
3{a_1}({a_0}^2 + {a_1}^2{k_p}) + (9{a_0}{a_1}^2 + 15{a_1}^2{k_p}){\omega _o}\nonumber \\ +& {a_1}(10{a_0} + 9{a_1}^2 + 18{k_p}){\omega _o}^2 + 30{a_1}^2{\omega _o}^3 + 33{a_1}{\omega _o}^4\nonumber \\ +& 8{\omega _o}^5 - 3{a_1}{a_0}{k_p} - 9{a_0}{k_p}\omega_o - 3{a_0}{\omega _o}^3
\label{eq-73}
\end{align}
	When $a_0 \ge 0$, $a_1 \ge 0$, and $\omega_o$ is sufficiently large, from (\ref{eq-73}), we have $B_0 > 0$, $B_1> 0$, $C_1> 0$, $C_2> 0$, $C_3 \ge 0$, and $C_4> 0$.
	According to Routh-Hurwitz criterion, if $B_0>0$, $B_1>0$, $C_1>0$, $C_2>0$, $C_3\ge0$, and $C_4>0$, then all the roots of ${}^2P(s) = 0$ are located in the left plane.

In summary, choosing ${\beta _i} = C_{n + 1}^i{\omega _o}^i$ for $i = 1,2,3$, when $m=n=2$, $a_1 \ge 0$ and $a_0 \ge 0$, there always exists a constant $\omega_o > 0$, such that the closed-loop is BIBO stable.
\eproof

\bibliographystyle{IEEEtran}
\bibliography{IEEEabrv,Bibliography}

\vfill


\end{document}